\documentclass[aps, amsmath, amssymb, amsfonts, twocolumn, superscriptaddress, footinbib]{revtex4}
\usepackage{mathrsfs}
\usepackage{amsthm}
\usepackage{graphicx}
\usepackage{subfigure} 
\newtheorem{thm}{Theorem}
\usepackage{float}
\usepackage{dcolumn}
\usepackage{bm}
\usepackage{bbold}
\usepackage{textcomp}
\usepackage{amsmath}
\usepackage[titletoc]{appendix}
\usepackage{tabularx}
\usepackage{verbatim}
\usepackage{booktabs}
\usepackage{algorithm,algorithmic}

\usepackage{hyperref}
\hypersetup{                    
	colorlinks,
	citecolor=blue,
	filecolor=blue,
	linkcolor=blue,
	urlcolor=blue
}

\begin{document}

	\title{Post-selection in noisy Gaussian boson sampling: part is better than whole}
	
	\author{Tian-Yu Yang}
	\affiliation{ State Key Laboratory of Low Dimensional Quantum Physics, Department of Physics, \\
		Tsinghua University, Beijing 100084, China}
	\author{Yi-Xin Shen}
	\affiliation{ State Key Laboratory of Low Dimensional Quantum Physics, Department of Physics, \\ Tsinghua University, Beijing 100084, China}
	\author{Zhou-Kai Cao}
	\affiliation{ State Key Laboratory of Low Dimensional Quantum Physics, Department of Physics, \\ Tsinghua University, Beijing 100084, China}
	\author{Xiang-Bin Wang}
	\email{ xbwang@mail.tsinghua.edu.cn}
	\affiliation{ State Key Laboratory of Low Dimensional Quantum Physics, Department of Physics, \\ Tsinghua University, Beijing 100084, China}
	\affiliation{ Jinan Institute of Quantum technology, SAICT, Jinan 250101, China}
	\affiliation{ Shanghai Branch, CAS Center for Excellence and Synergetic Innovation Center in Quantum Information and Quantum Physics, University of Science and Technology of China, Shanghai 201315, China}
	\affiliation{ Shenzhen Institute for Quantum Science and Engineering, and Physics Department, Southern University of Science and Technology, Shenzhen 518055, China}
	\affiliation{ Frontier Science Center for Quantum Information, Beijing, China}
	\centerline{}
	
	\begin{abstract}
		Gaussian boson sampling is originally proposed to show quantum advantage with quantum linear optical elements.
		Recently, several experimental breakthroughs based on Gaussian boson sampling pointing to quantum computing supremacy have been presented.
		However, due to technical limitations, the outcomes of Gaussian boson sampling devices are influenced severely by photon loss. 
		Here, we present an efficient and practical method to reduce the negative effect caused by photon loss. With no hardware modifications, our method takes the data post-selection process that discards low-quality data according to our criterion to improve the performance of the final computational results, say part is better than whole. 
		As an example, we show that the post-selection method can turn a GBS experiment that would otherwise fail in a ``non-classical test" into one that can pass that test.
		Besides improving the robustness of computation results of current GBS devices, this photon loss mitigation method may also benefit the further development of GBS-based quantum algorithms.		
	\end{abstract}

	\maketitle	
	

	\section{Introduction}
	
	The potential speedup~\cite{nielsen_quantum_2010,preskill_quantum_2018,shor_algorithms_1994} of quantum algorithms over their classical counterparts makes quantum computation a hot topic nowadays.
	In the past few years, people have witnessed the fast development of quantum computing technologies.
	One of the central issues these years is to demonstrate the quantum supremacy. 
	Towards this end, several breakthroughs of quantum computing occurred in different physical systems, such as superconducting systems~\cite{arute_quantum_2019,wu_strong_2021,huang_superconducting_2020,gong_quantum_2021}, trapped ions~\cite{bruzewicz_trapped-ion_2019}, and quantum optics~\cite{zhong_quantum_2020,arrazola_quantum_2021,zhong_phase-programmable_2021,slussarenko_photonic_2019,bourassa_blueprint_2021}, etc.

	Notwithstanding these achievements, we are still in an era~\cite{preskill_quantum_2018} of noisy intermediate-scale quantum (NISQ).
	There are mainly two kinds of quantum algorithms that might show an advantage over their classical counterparts on NISQ devices. They are approximate optimization algorithms, e.g., variational quantum eigenvalue solver~\cite{McClean_2016}, quantum approximate optimization algorithm (QAOA)~\cite{farhi_quantum_2014}, and quantum sampling problems, e.g., random circuit sampling~\cite{bouland_complexity_2019,arute_quantum_2019,wu_strong_2021} and Gaussian boson sampling (GBS)~\cite{rahimi-keshari_what_2015,hamilton_gaussian_2017,kruse_detailed_2019,quesada_gaussian_2018,zhong_quantum_2020,arrazola_quantum_2021,zhong_phase-programmable_2021,madsen_quantum_2022}. Due to the equipment imperfections on NISQ devices, errors frequently occur while implementing those quantum algorithms. Besides, quantum error correction is very expensive in NISQ devices, as they only own tens of qubits and high-error quantum gates. Thus, it will be essential to implement error mitigation methods to improve the performance of NISQ devices.

	The GBS problem is a variant of the boson sampling problem proposed by Aaronson and Arkhipov~\cite{aaronson_computational_2010}. GBS was originally designed to facilitate implementation of the original boson sampling problem without changing the problem’s computation complexity.
	Besides showing evidence of quantum advantage, GBS devices have been linked to several potential applications, such as quantum chemistry~\cite{huh_vibronic_2017,banchi_molecular_2020}, graph theory~\cite{bradler_gaussian_2018,arrazola_using_2018,bradler_graph_2021}, quantum approximate optimization~\cite{arrazola_quantum_2018}, and quantum machine learning~\cite{schuld_quantum_2019}.

	Although several applications based on GBS have been proposed, little about errors and noise in real GBS devices have been considered. In this work, we focus on error mitigation in the GBS problem.  In a GBS device, one of the main error sources is photon loss. Recently, error mitigation of photon loss in GBS devices has been studied in~\cite{su_error_2021}, which presents two schemes to mitigate the effect of photon loss based on extrapolation technique or probability estimation. 
	For the first time, these two schemes show that error mitigation methods are applicable for computing models based on quantum photonic devices.
	However, the methods in~\cite{su_error_2021} do not work in the case of large photon loss and hence the expected quantum advantage of the result could be undermined. 
	In addition, the existing error mitigation methods aim to recover the output probability of GBS devices. In practice, it requires collecting a sufficient number of samples to obtain the exact probability distribution satisfied by the sample.

	Here we present an efficient method to mitigate photon loss in GBS devices with no need for any hardware modification. The main idea of our method is to take a classical data post-selection process, which follows an insight into the similarity among different sets of input quantum states. 
	In comparison with methods in~\cite{su_error_2021}, our method is more robust to errors caused by photon loss. Moreover, our method does not need to collect the samples from the quantum computing devices for probability distribution of output patterns, say, it generates high-quality samples directly.

	We also analyze how the post-selection method will enhance the quantum advantage results of GBS experiments. Taking a non-classical test method~\cite{qi_regimes_2020} that is widely used by existing GBS experiments~\cite{arrazola_quantum_2021,zhong_phase-programmable_2021,madsen_quantum_2022} as an example, our post-selection method can make a GBS experiment that would have been judged by the method to have no quantum advantage become potentially quantum advantageous. That is, a GBS experiment that could have been approximately simulated by a classical algorithm becomes impossible to be simulated by that classical algorithm after post-selection. This example, i.e., the post-selection method can help the GBS experiment to surpass the non-classical test, demonstrates the effectiveness of the post-selection method in enhancing the quantum advantageous results of the GBS experiment.
	
	The post-selection method can also help to increase the circuit depth of GBS devices. Shallow circuit depth is another issue that affects the quantum advantage of existing GBS experiments~\cite{oh_classical_2022}. The circuit depth of the current GBS experiments is generally shallow, mainly due to the fact that the photon loss of GBS devices increases exponentially with increasing circuit depth. Post-selection method, which performs well at any photon loss rate, can help experimenters mitigate the increase in photon loss while increasing the circuit depth.

	This article is organized as follows: In section~\ref{sec:2} We will give an overview of the GBS protocol and provide some basic knowledge about quantum optics and the Gaussian state. 
	In section~\ref{sec:3}, we will prove several theorems which will demonstrate the correctness of the post-selection method. In section~\ref{sec:4}, we explain the post-selection method in detail and analyze the effect of the post-selection method on photon loss mitigation through numerical simulations. We also compare the differences between the post-selection method and the existing methods. In section~\ref{sec:5}, we analyze the effect of the post-selection process  on some other errors commonly found in GBS experiments including non-uniform loss, the limited photon number resolution (PNR) capability of the detector, and the dark count rate. In section~\ref{sec:6}, we show the role of post-selection method in enhancing the quantum advantage of GBS experimental results and the help of post-selection method in increasing the circuit depth of GBS devices.
	A summary is presented in section \ref{sec:7}.

	\section{Gaussian boson sampling}\label{sec:2}
\begin{figure}
		\centering
		\includegraphics[width=0.8\linewidth]{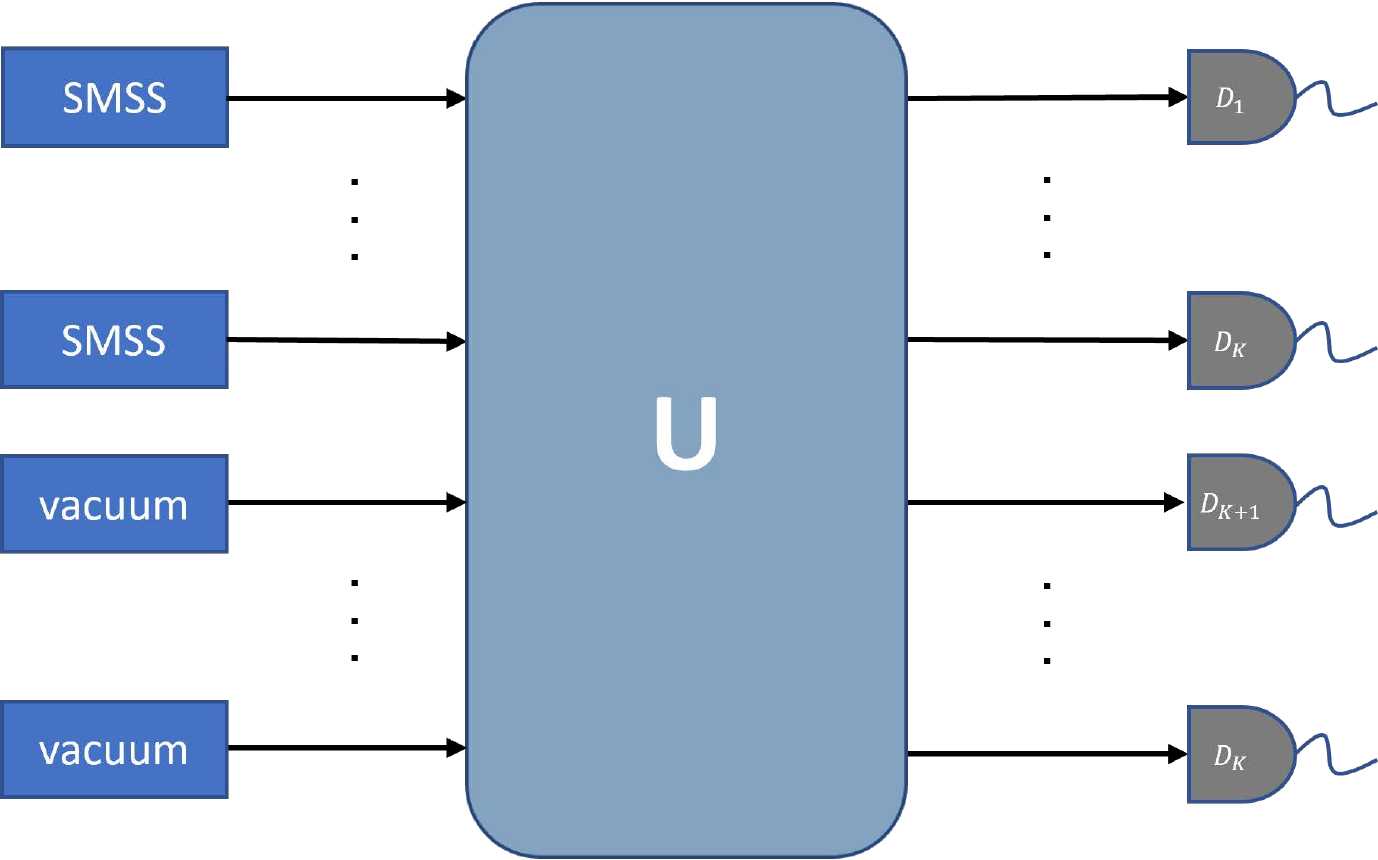}
		\caption{Schematic setup of Gaussian boson sampling: First, $K$ single-mode squeezed states (SMSSs) are injected into a passive linear optical network with $M$ modes. Then the output quantum state is measured by photon number resolving detectors $D_1 , D_2 , ... , D_M$ in each mode and an output pattern $\bar{n}=n_1 n_2 ... n_M$ is generated.}
		\label{fig:1}
	\end{figure}

	The architecture of the GBS~\cite{kruse_detailed_2019} is shown in Fig \ref{fig:1}.
	First, an input quantum state consists of $K$ single-mode squeezed states (SMSSs), and $M-K$ vacuum states are injected into an $M$-mode passive linear optical network. Photon number resolving detectors then measure the output quantum state, and an output pattern $\bar{n}=n_1n_2...n_M$, where $n_i$ is the detected photon number in the $i\text{th}$ mode, is generated. The probability distribution of an output pattern $\bar{n}=n_\text{1}n_\text{2}...n_\text{M}$ is~\cite{kruse_detailed_2019}
	\begin{equation}
		P_{\text{out}}(\bar{n})=\frac{1}{\bar{n}! \sqrt{\vert \sigma_Q \vert }} \mathrm{Haf} (\text{A}_S)
		,\label{eq:7}
	\end{equation}
	where $\mathrm{Haf}$ is a computationally hard matrix function called hafnian, $\sigma_Q = \sigma + \frac{I_{\text{2M}}}{2}$, $\text{A}_S$ is a matrix determined by GBS inputs and the passive linear optical network, and $\sigma$ is the covariance matrix of the output Gaussian state. 
	
	As shown in~\cite{wang_quantum_2007,serafini_quantum_2017,barnett_methods_2002,scully_quantum_1997}, SMSSs can be described as follows:
	\begin{small}
		\begin{equation}\label{eq:19}
			\begin{aligned}
				\vert r,\phi\rangle &=\hat{S}(r,\phi)|0\rangle\\
				&=\frac{1}{\sqrt{\cosh(r)}}\exp(-\frac{1}{2}\exp(i\phi)(\hat{a}^{\dagger})^2\tanh r)|0\rangle\\
				&=\frac{1}{\sqrt{\cosh(r)}}\sum_{n=0}^{\infty}\frac{\sqrt{(2n)!}}{n!}(-\frac{1}{2}\exp(i\phi)\tanh r)^n|2n\rangle,\\
			\end{aligned}
		\end{equation}
	\end{small}where $\hat{S}(r,\phi)$ is the squeezing operator, $r$ is squeezing strength and $\phi$ is the phase. 
	The photon number distribution of SMSS is 
	\begin{equation}\label{eq:3}
		p^{(r)}(n) =\begin{cases}
			\frac{1}{\cosh r}\frac{(n)!}{2^{n}((n/2)!)^2}(\tanh r)^{n} & n =0,2,4,... \\
			0 & n =1,3,5,...\\
		\end{cases} ,
	\end{equation}
	which corresponds to the negative binomial distribution~\cite{rohatgi_introduction_2015}.
	
	The total photon number distribution for the injected quantum state can be calculated through the convolution of probability distributions of single SMSSs. 
	The total photon number distribution for input SMSSs is thus
	\begin{equation}
		\begin{aligned}
			P^{\{r_i\}}(|\bar{n}|) = \sum_{n_1+...+n_K=N} \prod_{i=1}^{K} \binom{\frac{1}{2} +\frac{n_i}{2}-1}{\frac{n_i}{2}} \frac{\left(\tanh r_i \right)^{n_i} }{\cosh r_i},
			\label{eq:23}
		\end{aligned}
	\end{equation}
	where ${r_i}$ represents the squeezing strength of mode $i$ of input SMSSs, $\bar{n} = n_1n_2 \dots n_K$, $|\bar{n}|=\sum_{i=1}^K n_i$ and $n_i$ is even number represents the photon number in mode $i$.
	When each mode of input SMSSs has the same squeezing strength $r$, i.e., $r_i = r$ for $i=1,2,...,K$, the probability distribution of the total photon number is
	\begin{equation}
		\begin{aligned}
			P^{(r)}(N)&=\tbinom{N/2+K/2-1}{N/2} (\frac{1}{\cosh(r)})^K \tanh ^{N} (r)\\ \label{eq:5}
		\end{aligned}
	\end{equation}
	for $N=0,2,4,...$ and
	\begin{equation}
		P^{(r)}(N)=0,
	\end{equation}
	for $N=1,3,5,...$. Here $N$ is the total photon number. The mode (most common number) of this distribution is 
	\begin{equation}
		n_0=2\lfloor(K/2-1)\sinh^2(r) \rfloor .\label{eq:4}
	\end{equation}
	
	Passive linear optical networks preserve the photon number corresponding to the quantum states of the input and output.
	Uniform photon loss in passive linear optical network can be described as a larger passive linear optical network, including actual modes and loss (environment) modes~\cite{brod_classical_2020,garcia-patron_simulating_2019,oszmaniec_classical_2018}. Denote unitary transform corresponding to a passive linear optical network as  $\hat{U_0}$ and $\hat{\boldsymbol{a}}^\dagger=(\hat{a}_1^\dagger,\hat{a}_2^\dagger,...,\hat{a}_{2M}^\dagger)^T$, with $\hat{a}_i^\dagger$ the creation operator.
	The first $M$ modes are actual modes, and the last $M$ modes are loss (environment) modes initially in the vacuum state and will be traced out.
	When photon loss is uniform, we have
	\begin{equation}
		\hat{U}_0 \hat{\boldsymbol{a}}^\dagger \hat{U}_0^\dagger=\Lambda_0 \hat{\boldsymbol{a}}^\dagger,
		\label{eq:32}
	\end{equation}
	where 
	\begin{equation}
		\Lambda_0=\left(
		\begin{array} {cc}
			\oplus_{i=1}^M\sqrt{\eta}&\oplus_{i=1}^M\sqrt{1-\eta}\\
			-\oplus_{i=1}^M\sqrt{1-\eta}&\oplus_{i=1}^M\sqrt{\eta}\\
		\end{array}
		\right)
		,
	\end{equation}
	and $0\le\eta\le 1$ is the transmission rate.
	Moreover, denote the unitary transform corresponding to the passive linear optical network in GBS as $\hat{U_1}$. 
	We have
	\begin{equation}
		\hat{U}_1 \hat{\boldsymbol{a}}^\dagger \hat{U}_1^\dagger=\Lambda_1 \hat{\boldsymbol{a}}^\dagger,
		\label{eq:33}
	\end{equation}
	where 
	\begin{equation}
		\Lambda_1=\left(
		\begin{matrix} 
			\Lambda_M&0\\
			0&I_M\\
		\end{matrix}
		\right),
	\end{equation}
	$\Lambda_M$ is a $M\times M$ unitary matrix and $I_M$ is a identity matrix with dimension $M$.
	Considering a lossy GBS experiment, the output state is thus 
	\begin{equation}
		\hat{\rho}_{\text{out} }= \mathrm{Tr}_{M} \left( \hat{U_1}\hat{U_0}(\hat{\rho}_{\text{in}}\otimes\left|\bar{0}\right\rangle\left\langle\bar{0}\right|) \hat{U_0}^{\dagger}\hat{U_1}^{\dagger} \right).
		\label{eq:14}
	\end{equation}

	A brief introduction to photon detectors can be found in a recent article~\cite{slussarenko_photonic_2019}. There are three main errors in photon detectors, including detection efficiency, PNR capability and dark count rate. In theoretical analysis, the detection efficiency of a detector can be considered as a part of the total transmittance. PNR capability indicates the upper limit of the number of photons that can be accurately resolved by the detector. The dark count rate indicates that the detector has a certain probability of counting the cases where no photons reach the detector.

	\section{post-selection in Gaussian boson sampling}\label{sec:3}
	\subsection{Probability of different output pattern}\label{sec:31}
	Here we derive the expression of the probability distribution of GBS outputs in a different form from Eq.~(\ref{eq:7}), which will be very useful when proving our error mitigation method later in subsection \ref{sec:4}. When there is no photon loss, the input state is a pure state
	\begin{equation}
		\hat{\rho}_{\text{in}} = \otimes_{i=1}^M\left(\vert\xi_i,\phi_i\rangle \langle \xi_i, \phi_i \vert\right).
	\end{equation}
	For simplicity and without loss of generality, we can set the phase $\phi = \pi$. This is because the change in the phase of the input squeezed state is equivalent to the change in the phase of the passive linear optical network.
	The probability of output pattern can be expressed as
	\begin{equation}
		\begin{aligned}
			P_{\text{out}}(\bar{m})&=
			\langle\bar{m}|\hat{U}\hat{\rho}_{\text{in}}\hat{U}^\dagger|\bar{m}\rangle\\
			&=\sum_{\bar{n},\bar{l}}\langle\bar{m}|\hat{U}|\bar{n}\rangle\langle \bar{n}|\hat{\rho}_{\text{in}}|\bar{l}\rangle \langle \bar{l}|\hat{U}^\dagger|\bar{m}\rangle
		\end{aligned}.
	\end{equation}
	Denote $\sqrt{P_{\text{in}}(\bar{n})}=\langle\bar{n}|\xi\rangle=\sqrt{\langle\bar{n}| \hat{\rho}_{\text{in}}  | \bar{n}\rangle}$.
	We get
	\begin{equation}
		P_{\text{out}}(\bar{m})
		=\sum_{\bar{n},\bar{l}} v(\bar{m}|\bar{n})v^*(\bar{m}|\bar{l}) \sqrt{P_{\text{in}}(\bar{n})P_{\text{in}}(\bar{l})}
		,
		\label{eq:22}
	\end{equation}
	where $v(\bar{m}|\bar{n})=\langle\bar{m}|\hat{U}|\bar{n}\rangle$. As passive linear optical networks preserve the photon number of quantum optical states, we have
	\begin{equation}
		v(\bar{m}|\bar{n})=0,
	\end{equation}
	when $\sum_{i=0}^M m_i \ne \sum_{i=0}^M n_i$.
	
	Next, considering the situation when photon loss exists. According to Eq.~(\ref{eq:14})
	\begin{equation}
		\begin{aligned}
			P_{\text{out}}(\bar{m})&=
			\langle\bar{m}|\mathrm{Tr}_{M} \left( \hat{U}_1\hat{U}_0
			(\hat{\rho}_{\text{in}}\otimes|\bar{0}\rangle\langle\bar{0}|)
			\hat{U}_0^{\dagger}\hat{U}_1^{\dagger} \right)|\bar{m}\rangle\\
			&=\sum_{\bar{n},\bar{l}} v_1(\bar{m}|\bar{n})v_1^*(\bar{m}|\bar{l})\sum_{\bar{k}}\sum_{\bar{i},\bar{j}}
			\langle \bar{n},\bar{k}|\hat{U}_0|\bar{i},\bar{0}\rangle\\
			&\quad\times\langle\bar{i}|\hat{\rho}_{\text{in}}|\bar{j}\rangle\langle\bar{j},\bar{0}|\hat{U}_0^\dagger|\bar{l},\bar{k}\rangle,
			\label{eq:21}
		\end{aligned}
	\end{equation}
	where $\hat{U}_1$ and $\hat{U}_0$ are given in Eq.~(\ref{eq:32}) and (\ref{eq:33}), $v_1(\bar{m}|\bar{n})=\langle\bar{m}|\hat{U}_1|\bar{n}\rangle$ and $v^*_1(\bar{m}\mid\bar{l})=\langle\bar{l}|\hat{U}_1^\dagger|\bar{m}\rangle$.
	According to Eq.~(\ref{eq:32}), we have
	\begin{small}
		\begin{equation}
			\hat{U}_0|\bar{i},\bar{0}\rangle
			=\frac{1}{\sqrt{(i_1!)(i_2!)...(i_{M}!)}}
			\prod_{p=1}^M\left(\sqrt{\eta}\hat{a}_p^\dagger + \sqrt{1-\eta}\hat{a}_{p+M}^\dagger \right)^{i_p}\left|\bar{0} ,\bar{0}\right\rangle.
		\end{equation}
	\end{small}
	Then we have
	\begin{small}
		\begin{equation}
			\begin{aligned}
				\langle\bar{n},\bar{k}|\hat{U}_0|\bar{i},\bar{0}\rangle=&
				\langle\bar{n},\bar{k}|\frac{1}{\sqrt{(i_1!)(i_2!)...(i_{M}!)}}\\
				&\times \prod_{p=1}^M\binom{i_p}{n_p}
				\left( \sqrt{\eta}\hat{a}_p^\dagger \right)^{n_p}
				\left( \sqrt{1-\eta}\hat{a}_{p+M}^\dagger \right)^{i_p-n_p}\left|\bar{0} ,\bar{0}\right\rangle
				.
			\end{aligned}
		\end{equation}
	\end{small}
	This equation is not zero only when 
	\begin{equation}
		\bar{k}=\bar{i}-\bar{n}.
	\end{equation}
	So, we get
	\begin{equation}
		\langle\bar{n},\bar{k}|\hat{U}_0|\bar{i},\bar{0}\rangle
		=\prod_{p=1}^M \sqrt{\binom{i_p}{n_p}} \left(\sqrt{\eta}\right)^{n_p}
		\left(\sqrt{1-\eta}\right)^{i_p-n_p}
		.
		\label{eq:20}
	\end{equation}
	Finally, bring Eq.~(\ref{eq:20}) back to Eq.~(\ref{eq:21}), we get
	\begin{small}
		\begin{equation}
			\begin{aligned}
				P_{\text{out}}(\bar{m})=&\sum_{\bar{n},\bar{l}}v_1(\bar{m}|\bar{n})v_1^*(\bar{m}|\bar{l})\sum_{\bar{i}\ge\bar{n}}\sum_{\bar{j}\ge\bar{l}}\prod_{p=1}^{M}\prod_{q=1}^{M}\sqrt{P_{\text{in}}(\bar{i})P_{\text{in}}(\bar{j})}\\
				&\times\quad\sqrt{\binom{\bar{i}}{\bar{n}}\eta^{n_p}(1-\eta)^{i_p-n_p}} 
				\sqrt{\binom{\bar{j}}{\bar{l}}\eta^{l_q}(1-\eta)^{j_q-l_q}} \\
				=&\sum_{\bar{n},\bar{l}}v_1(\bar{m}|\bar{n})v_1^*(\bar{m}|\bar{l}) \sqrt{P^{(\eta)}_{\text{in}}(\bar{n})P^{(\eta)}_{\text{in}}(\bar{l})},
				\label{eq:1}
			\end{aligned}
		\end{equation}
	\end{small} where 
	\begin{equation}
		P^{(\eta)}_{\text{in}}(\bar{n})=\sum_{\bar{i}\ge\bar{n}} \prod_{p=1}^{M}\binom{\bar{i}}{\bar{n}}\eta^{n_p}(1-\eta)^{i_p-n_p}P_{\text{in}}(\bar{i}),
	\end{equation}
	and
	\begin{equation}
		\binom{\bar{i}}{\bar{n}} = \binom{i_1}{n_1}\binom{i_2}{n_2}...\binom{i_M}{n_M}.
	\end{equation}

	\subsection{Mapping between different GBS experiments}\label{sec:32}
	Firstly, we consider the simplest case, i.e., the case when there is no photon loss. We first give Theorem~\ref{thm:1}. Theorem~\ref{thm:1} shows that GBS experiments with different input squeezing strengths can be mapped to each other under certain conditions.
	This result inspires us to find an efficient and practical photon loss mitigation method. 
	\begin{thm}
		Denote the probability distribution of output patterns of GBS devices with input ${\left\{r_i\right\}}$ as $\mathrm{P}^{\left\{r_i\right\}}\left(\bar{n}\right)$, where ${\left\{r_i\right\}}=\left\{r_1,r_2,...,r_{M} \right\}$ represents the input squeezed states with the squeezing parameters ${\left\{r_i\right\}}$, and $\bar{n}=n_1...n_{M}$ is the output pattern. We have 
		\begin{equation}
			P_{\text{out}}^{\{r^{\prime}_i\}}\left(\bar{n}\right) = P_{\text{out}}^{\{r_i\}}\left(\bar{n}\right) \times \frac{P_{\text{out}}^{\{r^{\prime}_i\}}\left(\left|\bar{n}\right|\right)}{P_{\text{out}}^{\{r_i\}}\left(\left|\bar{n}\right|\right)},
		\end{equation}
		provided that $\left\{r^{\prime}_{i}\right\}$ satisfies $\frac{\tanh{r_i}}{\tanh{r^{\prime}_i}}=c$, for $i=1,2,...,M$ and $c$ is a constant real number.
		Here $\left|\bar{n}\right|=\sum_{n=1}^{\mathrm{M}}n_i$ is the total photon number, $n_i$ is even for all $i$ and $P^{\{r_i\}}\left(\left|\bar{n}\right|\right)$ is the probability of an output pattern $\bar{n}$ with a total photon number $\left|\bar{n}\right|$.
		\label{thm:1}
	\end{thm}

	\begin{proof}
		According to Eq.~(\ref{eq:3}), we have
		\begin{equation}
			\begin{aligned}
				\frac {P^{\{r_i\}}_{\text{in}} (\bar{n})}{P^{\{r^{\prime}_i\}}_{\text{in}} (\bar{n})}=\prod_{i=1}^{K} \frac{\cosh r^{\prime}_i}{\cosh r_i} \left( \frac{\tanh r_i}{ \tanh r^{\prime}_i} \right)^{n_i},\label{eq:9}
			\end{aligned}
		\end{equation}
		where $P^{\{r_i\}}_{\text{in}} (\bar{n})$ (or $P^{\{r^{\prime}_i\}}_{\text{in}} (\bar{n})$) are probability distribution of output pattern with input squeezing strength $\{r_i\}$ (or $\{r^{\prime}_i\}$), $\bar{n}=n_1 n_2 ... n_{K}$ and $\bar{m}=m_1 m_2 ... m_K$ are arbitrary input patterns with the same total photon number $N$, $n_i$ and $m_i$ ($i=1,...,{K}$) are photon numbers in each input mode. 
		If we choose $\{r'_i\}$ that satisfies $\frac{\tanh r_i}{ \tanh r'_i} =c$ for all $i=1,2,...,K$, we will find
		\begin{equation}
			\frac {P^{\{r_i\}}_{\text{in}} (\bar{n})}{P^{\{r^{\prime}_i\}}_{\text{in}} (\bar{n})}=\frac {P^{\{r_i\}}_{\text{in}} (\bar{m})}{P^{\{r^{\prime}_i\}}_{\text{in}} (\bar{m})}
			\label{eq:prob2}.
		\end{equation}
		
		According to Eq.~(\ref{eq:22}), we find
		\begin{equation}
			\begin{aligned}
				\frac{P_{\text{out}}^{\{r_i\}} (\bar{q})}{P_{\text{out}}^{\{r_i\}} (\bar{j})}&=\frac{\sum_{\bar{n},\bar{l}} v(\bar{q} \vert \bar{n}) v^{*}(\bar{q} \vert \bar{l}) \sqrt{P_{\text{in}}^{\{r_i\}}(\bar{n}) P_{\text{in}}^{\{r_i\}}(\bar{l})}}{\sum_{\bar{n}^{\prime},\bar{l}^{\prime}} v(\bar{j} \vert \bar{n}^{\prime}) v^{*}(\bar{j} \vert \bar{l}^{\prime}) \sqrt{P_{\text{in}}^{\{r_i\}}(\bar{n}^{\prime}) P_{\text{in}}^{\{r_i\}}(\bar{l}^{\prime})}}\\		
				\label{eq:prob3}
			\end{aligned},
		\end{equation}
		where $P^{\{r_i\}}_{\text{out}} (\bar{q})$ and $P^{\{r_i\}}_{\text{out}} (\bar{j})$ are probability distribution of output patterns with input squeezing strength $\{r_i\}$, $\bar{q}=q_1 q_2 ... q_K$ and $\bar{j}=j_1 j_2 ... j_K$ are arbitrary output patterns with the same total photon number $N$, $q_i$ and $j_i$ ($i=1,...,K$) are photon numbers in each output mode. 
		Plugging (\ref{eq:prob2}) into (\ref{eq:prob3}) and we get
		\begin{equation}
			\begin{aligned}
				\frac{P_{\text{out}}^{\{r_i\}} (\bar{q})}{P_{\text{out}}^{\{r_i\}} (\bar{j})}&=\sum_{\bar{n},\bar{l}}\frac{v(\bar{q} \vert \bar{n}) v^{*}(\bar{q} \vert \bar{l})}{\sum_{\bar{n}^{\prime},\bar{l}^{\prime}} v(\bar{j} \vert \bar{n}^{\prime}) v^{*}(\bar{j} \vert \bar{l}^{\prime}) \sqrt{\frac{P^{(r^{\prime}_i)}_{\text{in}}(\bar{n}^{\prime}) P^{\{r^{\prime}_i\}}_{\text{in}}(\bar{l}^{\prime})}{P^{\{r^{\prime}_i\}}_{\text{in}}(\bar{n}) P^{\{r^{\prime}_i\}}_{\text{in}}(\bar{l})}}}\\
				&=\frac{P_{\text{out}}^{\{r^{\prime}_i\}} (\bar{q})}{P_{\text{out}}^{\{r^{\prime}_i\}} (\bar{j})}.
			\end{aligned}
			\label{eq:28}
		\end{equation}
		Taking summation over $\bar{q}$, we obtain
		\begin{equation}
			\frac{P_{\text{out}}^{\{r_i\}}\left(\left|\bar{n}\right|\right)}{P_{\text{out}}^{\{r_i\}} (\bar{n})} = \frac{P_{\text{out}}^{\{r^{\prime}_i\}}\left(\left|\bar{n}\right|\right)}{P_{\text{out}}^{\{r^{\prime}_i\}} (\bar{n})}
		\end{equation}
		which completes the proof.
	\end{proof}

	We next give Theorem~\ref{thm:2}. Theorem~\ref{thm:2} shows that under uniform photon loss, GBS experiments with different input squeezing strength can still be mapped to each other under certain conditions, and the total transmission rate of the GBS device will vary in a certain proportion.

	\begin{thm}
		\label{thm:2}
		
		Denote the probability distribution of output patterns of GBS devices with input ${\left\{r_i\right\}}$ and overall transmission rate $\eta$ as $\mathrm{P}^{\left\{r_i,\eta\right\}}\left(\bar{n}\right)$, where ${\left\{r_i\right\}}=\left\{r_1,r_2,...,r_{M} \right\}$ represents the input squeezed states with the squeezing parameters ${\left\{r_i\right\}}$, and $\bar{n}=n_1...n_{M}$ is the output pattern. We have 
		\begin{equation}
			P_{\text{out}}^{\{r^{\prime}_i,\eta^{\prime}\}}\left(\bar{n}\right) = P_{\text{out}}^{\{r_i,\eta\}}\left(\bar{n}\right) \times \frac{P_{\text{out}}^{\{r^{\prime}_i\}}\left(\left|\bar{n}\right|\right)}{P_{\text{out}}^{\{r_i\}}\left(\left|\bar{n}\right|\right)}\left(\frac{\eta^{\prime}}{\eta}\right)^{\left|\bar{n}\right|},
		\end{equation}
		provided that $\left\{r^{\prime}_{i}\right\}$ satisfies $\frac{\tanh{r_i}}{\tanh{r^{\prime}_i}}=c$, for $i=1,2,...,M$ and $c$ is a constant real number.
		Here $\left|\bar{n}\right|=\sum_{n=1}^{\mathrm{M}}n_i$ is the total photon number and $P^{\{r_i\}}\left(\left|\bar{n}\right|\right)$ is the probability of an output pattern $\bar{n}$ with a total photon number $\left|\bar{n}\right|$, and $\eta^{\prime}=1-\left(1-\eta\right)c$.
	\end{thm}

	\begin{proof}
		
		According to Eq.~(\ref{eq:1}), we have
		\begin{equation}
			P_{\text{out}}^{\{r_i,\eta\}}\left(\bar{n}\right) = \sum_{\bar{m},\bar{l}}v_1(\bar{n}|\bar{m})v_1^*(\bar{n}|\bar{l}) \sqrt{P^{\{r_i,\eta\}}_{\text{in}}(\bar{m})P^{\{r_i,\eta\}}_{\text{in}}(\bar{l})},
			\label{eq:30}
		\end{equation}
		where
		\begin{equation}
			P^{\{r_i,\eta\}}_{\text{in}}(\bar{m})=\sum_{\bar{i}\ge\bar{m}} \prod_{p=1}^{M}\binom{\bar{i}}{\bar{m}}\eta^{m_p}(1-\eta)^{i_p-m_p}P^{\{r_i\}}_{\text{in}}(\bar{i}),
		\end{equation}
		and $P^{\{r_i\}}_{\text{in}}(\bar{i})$ is defined as in theorem \ref{thm:1}.
		It can be shown that
		\begin{equation}
			\begin{aligned}
				&P^{\{r_i,\eta\}}_{\text{in}}(\bar{m})\frac{P_{\text{out}}^{\{r^{\prime}_i\}}\left(\left|\bar{n}\right|\right)}{P_{\text{out}}^{\{r_i\}}\left(\left|\bar{n}\right|\right)}\left(\frac{\eta^{\prime}}{\eta}\right)^{\left|\bar{n}\right|}\\
				=&\sum_{\bar{i}\ge\bar{m}} \prod_{p=1}^{M}\binom{\bar{i}}{\bar{m}}\eta^{m_p}(1-\eta)^{i_p-m_p}P^{\{r_i\}}_{\text{in}}(\bar{i})\left(\frac{\eta^{\prime}}{\eta}\right)^{\left|\bar{n}\right|} \\
				&\times \prod_{q=1}^{K} \frac{\cosh r^{\prime}_q}{\cosh r_q} \left( \frac{\tanh r_q}{ \tanh r^{\prime}_q} \right)^{n_q}
				\\
				=&\sum_{\bar{i}\ge\bar{m}} \prod_{p=1}^{M}\binom{\bar{i}}{\bar{m}}\left(\eta'\right)^{m_p}\left(c\left(1-\eta\right)\right)^{i_p-m_p} \\
				&\times P^{\{r_i\}}_{\text{in}}(\bar{i})\left( \frac{\tanh r_p}{ \tanh r^{\prime}_p} \right)^{-i_p}\prod_{q=1}^{K} \frac{\cosh r^{\prime}_q}{\cosh r_q} 
				\\
				=&\sum_{\bar{i}\ge\bar{m}} \prod_{p=1}^{M}\binom{\bar{i}}{\bar{m}}{\eta'}^{m_p}(1-\eta')^{i_p-m_p}P^{\{r_i'\}}_{\text{in}}(\bar{i}) \\
				=&P^{\{r_i',\eta'\}}_{\text{in}}(\bar{m})
				,
				\label{eq:29}
			\end{aligned}
		\end{equation}
		where we used the relation $\eta^{\prime}=1-\left(1-\eta\right)c$.
		
		Combining Eq.~(\ref{eq:29}) and Eq.~(\ref{eq:30}), we can show the correctness of theorem \ref{thm:2}.
	\end{proof}

	Theorem~\ref{thm:2} shows that an error mitigation method for GBS devices by post-selection of the generated samples is possible, i.e., by post-selecting the samples output from the GBS devices according to a certain probability, it is possible to make the samples retained by post-selection equivalent to having a different transmission rate than the original output samples. We will discuss this post-selection scheme in detail in the next section.

	\section{photon loss mitigation}\label{sec:4}
	\subsection{photon loss mitigation protocol}\label{sec:41}
	Inspired by theorem 2, we propose the following post-selection method to mitigate the photon loss error in GBS experiments. For a sample obtained from a GBS experiment, we choose the following post-selection probability 
	\begin{equation}
		P_{\text{post}}(\vert \bar{n} \vert) = c ^{\alpha-\vert\bar{n}\vert} \left(\frac{\eta'}{\eta}\right)^{\vert \bar{n} \vert},\label{eq:25}
	\end{equation}
	where 
	\begin{align}
		\label{eq:41}
		\eta ' =1- \left(1-\eta\right) c,
	\end{align}
	and $c=\frac{\tanh r_i}{\tanh r'_i}<1$, to decide whether to keep the sample or not. 
	To ensure that the post-selection probability is less than $1$, we need to discard samples with a total photon number above a certain value. To be precise, we need to set a truncated photon number $N_0$. In the post-selection process, if the total number of photons of a sample exceeds the truncated photon number $N_0$, this sample needs to be discarded. At the same time, in order to retain as many samples as possible in the post-selection process, the parameter $\alpha$ in Eq.~(\ref{eq:25}) is chosen to satisfy the following equation
	\begin{align}
		P_{\text{post}}\left( N_0\right)=1,
		\label{eq:42}
	\end{align}
	This condition ensures that the probability of a sample with a total photon number less than $N_0$ being retained in post-selection is always less than $1$. This guarantees the soundness of the post-selection method.
	
	After a certain number of samples have been collected in an experiment, the experimenter can follow this post-selection probability to decide which of the samples will be retained and which will be discarded. Specifically, based on the total  photon number $\left|\bar{n}\right|$ in each sample $\bar{n}$, the experimenter uses Eq.~(\ref{eq:25}) to obtain the probability of retaining the sample. The experimenter then decides to keep or discard the sample based on this probability.  After this data post-processing process, the final retained samples are equivalently affected by less photon loss.
	Roughly speaking, we can simulate a GBS experiment with input squeezing strength $\left\{r^{\prime}_i\right\}$ and an overall transmission rate $\eta'$ by a GBS experiment with input squeezing strength $\left\{r_i\right\}$ and an overall transmission rate $\eta$.
	The above process is summarized in protocol~\ref{pro:1}. In the following we explain the details of this post-selection method.
	
	One might challenge whether our method really helps experimental physicists improve the quantum advantage of their GBS devices because of the existence of the truncated photon number. This questioning is mainly based on the fact that, usually, the quantum advantage of a GBS experiment is measured in terms of the maximum photon number of the detected samples. This is because for the GBS problem, the increase in the total photon number  obtained by sampling corresponds to an exponential increase in the time required by classical computers to simulate this sampling result~\cite{quesada_exact_2020}. Here we need to emphasize the point that the existence of a truncated photon number does not indicate that the post-selection method imposes a limit on the maximum number of photons that can be obtained in the experiment. That is, the worries that post-selection will limit the maximum photon number that can be obtained in the experiment are superfluous. In fact, in the experiment, the truncated photon number can be taken as the maximum total number of photons contained in a set of samples. According to the post-selection probability Eq.~(\ref{eq:25}) and (\ref{eq:42}) , samples with a total photon number equal to the truncated photon number will be retained $100\%$ of the time. This allows the post-selected samples to have the maximum total photon number. This method of selecting the number of truncated photons is advantageous for claiming a quantum advantage. Based on the above analysis, we emphasize here the fact that the post-selection method does not affect the difficulty of obtaining large photon number samples experimentally.
	
	Of course, depending on the experimenter's purpose, an arbitrary total photon number can be selected as the truncated photon number in the experiment. For example, when the experimenter wants to use the GBS sampling results to perform some GBS-related application problems, the truncated photon number can be set to such a value that corresponds to the total photon number most likely to occur in the experiment, e.g. mode number as given in Eq.~(\ref{eq:4}). This has the advantage that it can, to some extent, increase the proportion of the samples retained after the post-selection process. This is because when the photon number of a sample is smaller than the truncated photon number, the probability of the sample being retained decreases exponentially by the difference between its photon number and the truncated photon number. This is shown in the following equation:
	\begin{equation}
		\frac{P_{\text{post}}\left(|\bar{m}|\right)}{P_{\text{post}}\left(|\bar{n}|\right)}=\left(\frac{\eta'}{c\eta}\right)^{-\left(|\bar{n}|-|\bar{m}|\right)}.
	\end{equation}
	
	We can also calculate the yield of the post-selection process. By yield, we mean the percentage of the samples that are retained after the post-selection process. We calculate the yield in the asymptotic case without considering statistical fluctuations. The yield can be calculated using the following equation.
	\begin{equation}
		y_2 = \sum_{n=0}^{N_{0}} P_{\text{post}}(\vert \bar{n} \vert) P_{\text{out}}^{\{r_i,\eta\}}(\vert \bar{n} \vert) .
		\label{eq:31}
	\end{equation}
	From the above equation, it can be seen that the yield is affected by several experimental parameters, including the actual input squeezing strength $\{r_i\}$, the number of input squeezed states $K$, the overall transmission rate $\eta$, and the squeezing strength $\{r'_i\}$ of the input squeezed states in the GBS experiment to be simulated by the post-selection method.
	Although the post-selection method will reduce the total number of samples, it will give a mitigation of the sample error. We will demonstrate in the subsection that sacrificing a certain number of samples for a reduction in error is worthwhile for the demonstration of quantum advantage.

	\floatname{algorithm}{Protocol}
	\begin{algorithm}[!h]
		\caption{Photon loss mitigation}
		\normalsize
		\label{pro:1}
		\begin{algorithmic} 
			\STATE \textbf{Step 1}: Inject input SMSSs with parameter set $\{r_i\}$ into the passive linear optical network. The overall transmission rate is denoted as $\eta$. 
			\STATE \textbf{Step 2}: Measure the output state and calculate the total photon number.
			\STATE \textbf{Step 3}: According to the total photon number, we decide whether to retain or discard the specific outcome at that time based on post-selection probability given in Eq.~(\ref{eq:25}).
			\STATE \textbf{Step 4}: Repeat step $1-3$ for many times we obtain the outcomes with probability distribution given by the input SMSSs with parameter set $\{r_i'\}$ and overall transmission rate $\eta'$.
		\end{algorithmic}
	\end{algorithm}

	\subsection{analysis and numerical results}\label{sec:42}
	In this subsection, we will estimate the effect of the post-selection method on the improvement of the equivalent transmission rate $\eta'$ under different experimental conditions. 
	We will use the equations
	in~\ref{sec:41} to perform our simulation.
	Although the simulation here is based on the analytic equations in~\ref{sec:41} (i.e., only one source of error, photon loss in the GBS experiment, is considered), it provides a good demonstration of how the post-selection method performs under real-world conditions of use. The reason for this statement will be given in subsection~\ref{sec:5}, where we will analyze the effect of other possible errors on the post-selection process, which includes common sources of errors in GBS experiments, e.g., non-uniform loss, limited PNR capability and dark counts.

\begin{figure}[htbp]
		\centering
		\subfigure[Variation of $\eta'_{max}$ with $\eta$ for different squeezing strengths $r$]{\label{fig:21}
			\includegraphics[width=0.95\linewidth]{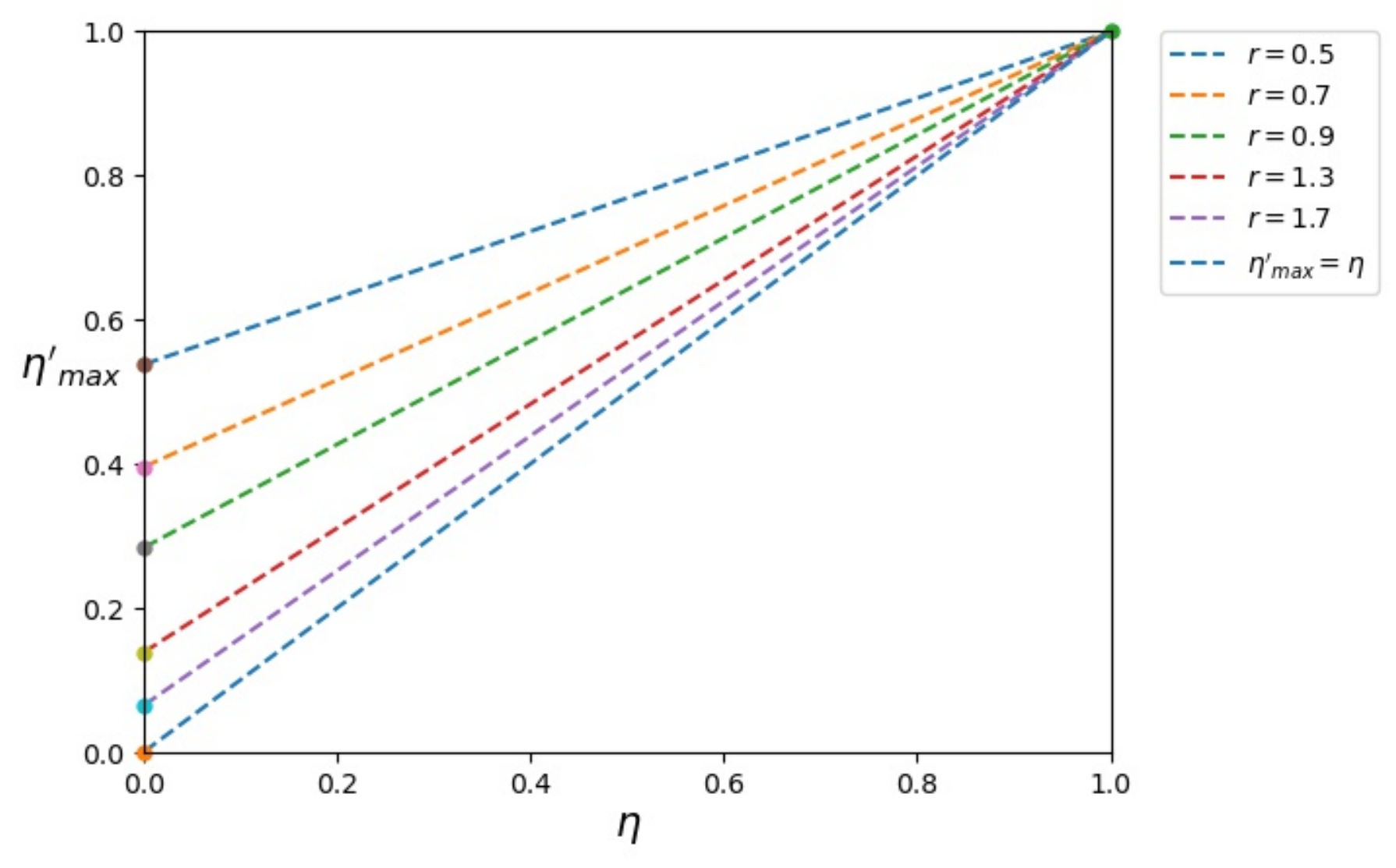}
		}
		\subfigure[Variation of $\eta'_{max}$ with $r$ for different transimission rate $\eta$]{\label{fig:22}
			\includegraphics[width=0.95\linewidth]{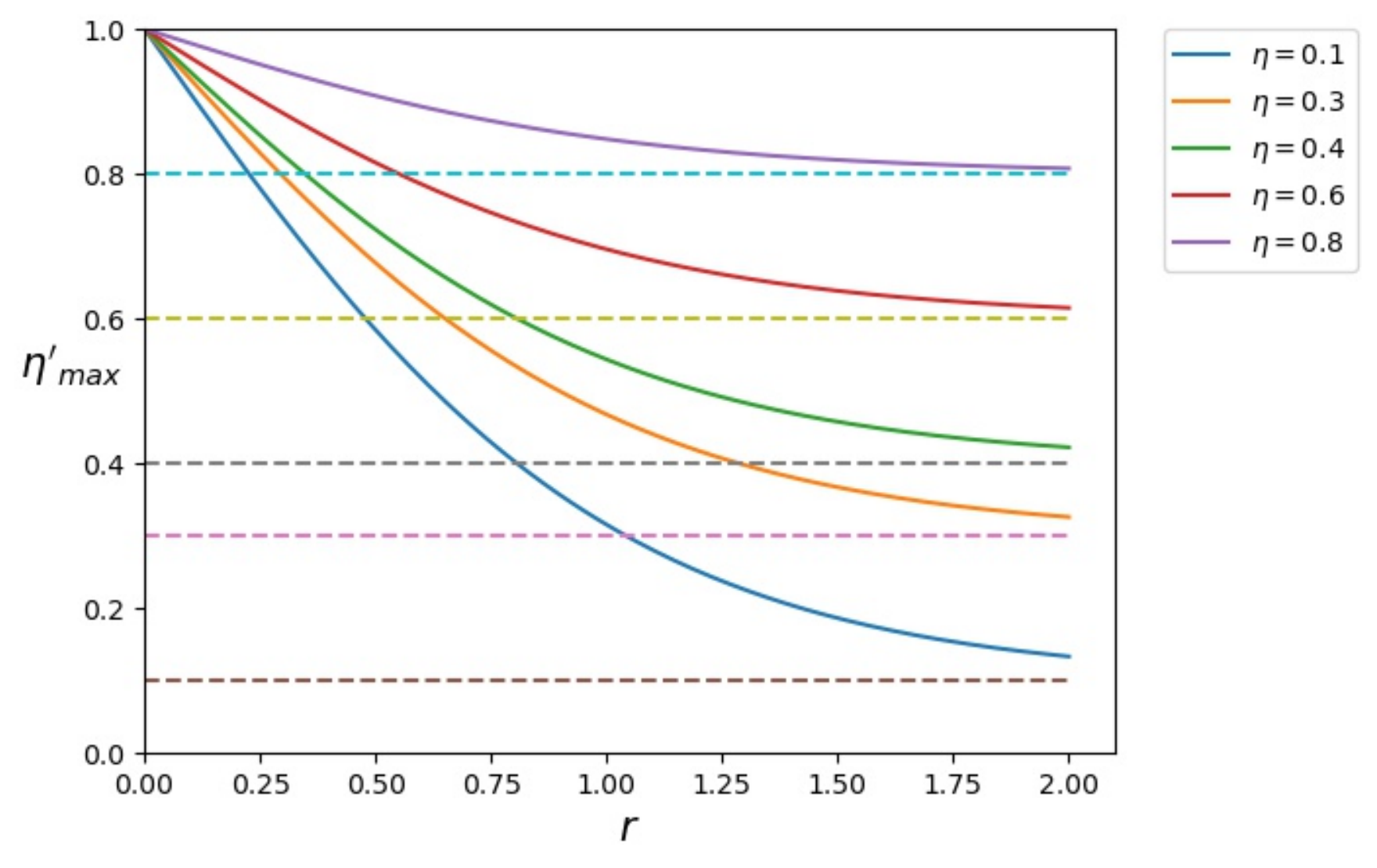} 
		}
		\DeclareGraphicsExtensions.
		\caption{\label{fig:2} Numerical simulations to show that photon loss mitigation method given in protocol 1 performs better for larger photon loss and smaller squeezing strength $r$. }
	\end{figure}

	In a GBS experiment, the actual input squeezing strength of the squeezing state imposes a limit on the ability of post-selection to reduce the effect of photon loss on the experimental results. This can be seen from Eq.~(\ref{eq:41}). Since $\tanh r'<1$, if the value of $r$ is taken as fixed, we find that
	\begin{align}
		\eta'=1-\left(1-\eta\right)\frac{\tanh r}{\tanh r'}<1-\left(1-\eta\right)\tanh r.
	\end{align}
	Therefore, the smaller the actual input squeezing strength on the experiment, the more the equivalent transmission rate increased by the post-selection process . Also, the smaller the transmission rate $\eta$, the larger the range of change $\Delta \eta = \eta'-\eta$ under post-selection. These two properties are shown visually in Fig~\ref{fig:2}. In Fig~\ref{fig:2}, $\eta'_{max}=1-\left(1-\eta\right)\tanh r$ indicates the maximum equivalent transmission rate that can be obtained by the post-selection method, and the other symbols have the same meaning as above. 
	Fig~\ref{fig:21} shows the relationship between the maximum equivalent transmission rate $\eta'_{max}$ and the actual transmission rate $\eta$ in the experiment for several squeezing strengths. It can be seen that as the squeezing strength $r$ increases, the enhancement of transmission rate by post-selection decreases.
	Fig~\ref{fig:22} shows the relationship between the maximum equivalent transmission rate $\eta'_{max}$ and the actual input squeezing strength for several actual transmission rates. It can be seen that as the actual transmission rate $\eta$ increases, the increase in transmission rate brought by the post-selection decreases. Although Fig~\ref{fig:22} shows that $\eta'_{max}$ becomes close to 1 as $r$ approaches 0, this does not mean that experimentally a very small squeezing strength $r$ should be used. This is because as the squeezing strength decreases, more squeezed states need to be input to obtain samples with a larger total photon number. We can take advantage of the properties of the post-selection method described above to help us to increase the circuit depth of GBS devices, which will be shown later in subsection~\ref{sec:62}.
	
\begin{figure}[htbp]
		\centering
		\subfigure[$r = 1.1$, $K = 216$, $\eta = 0.32$, $N_{0}=219 $]{\label{fig:31}
			\includegraphics[width=0.95\linewidth]{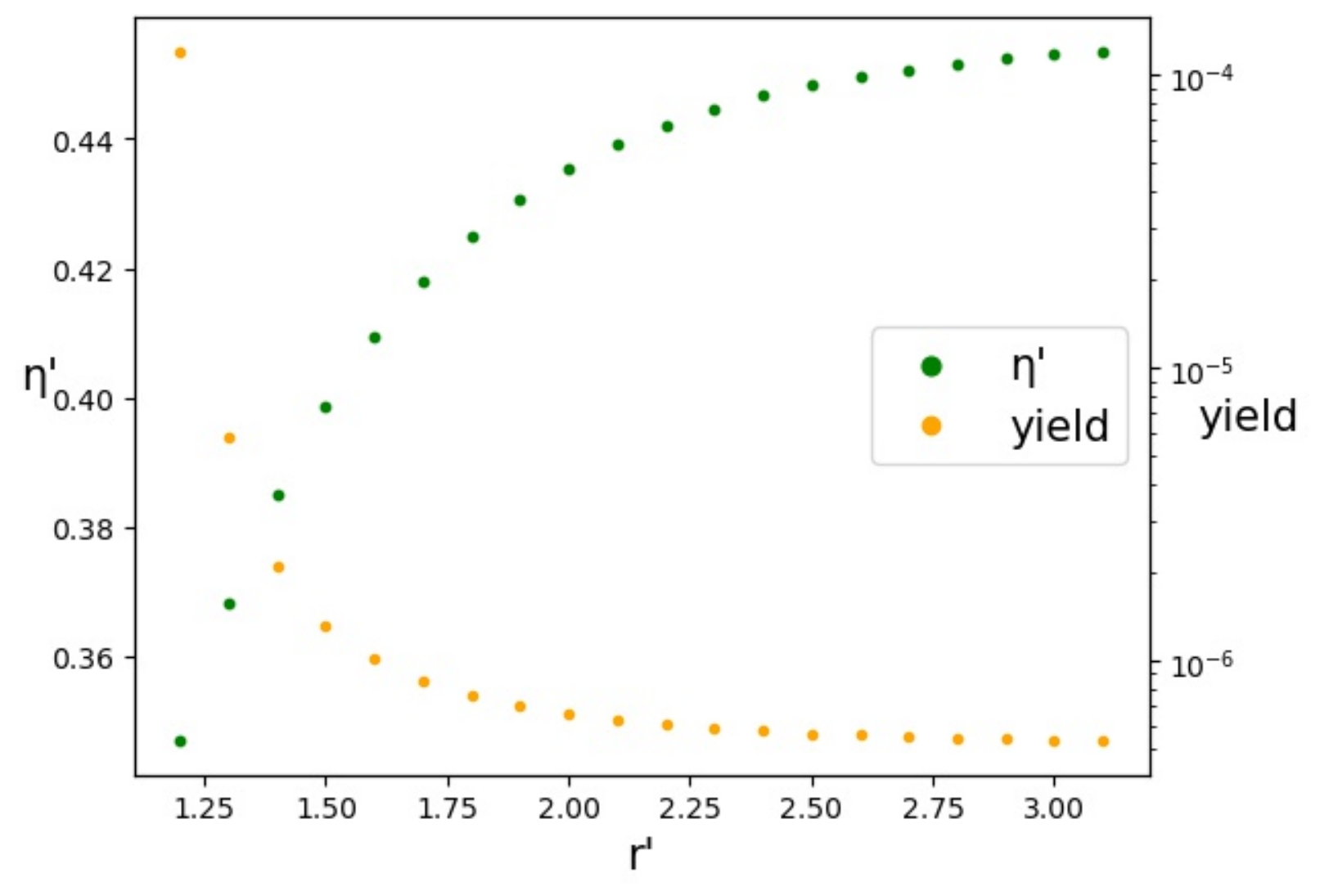}
		}
		\subfigure[ $r = 1.4$, $K = 50$, $\eta = 0.5$, $N_{0}=113 $]{\label{fig:32}
			\includegraphics[width=0.95\linewidth]{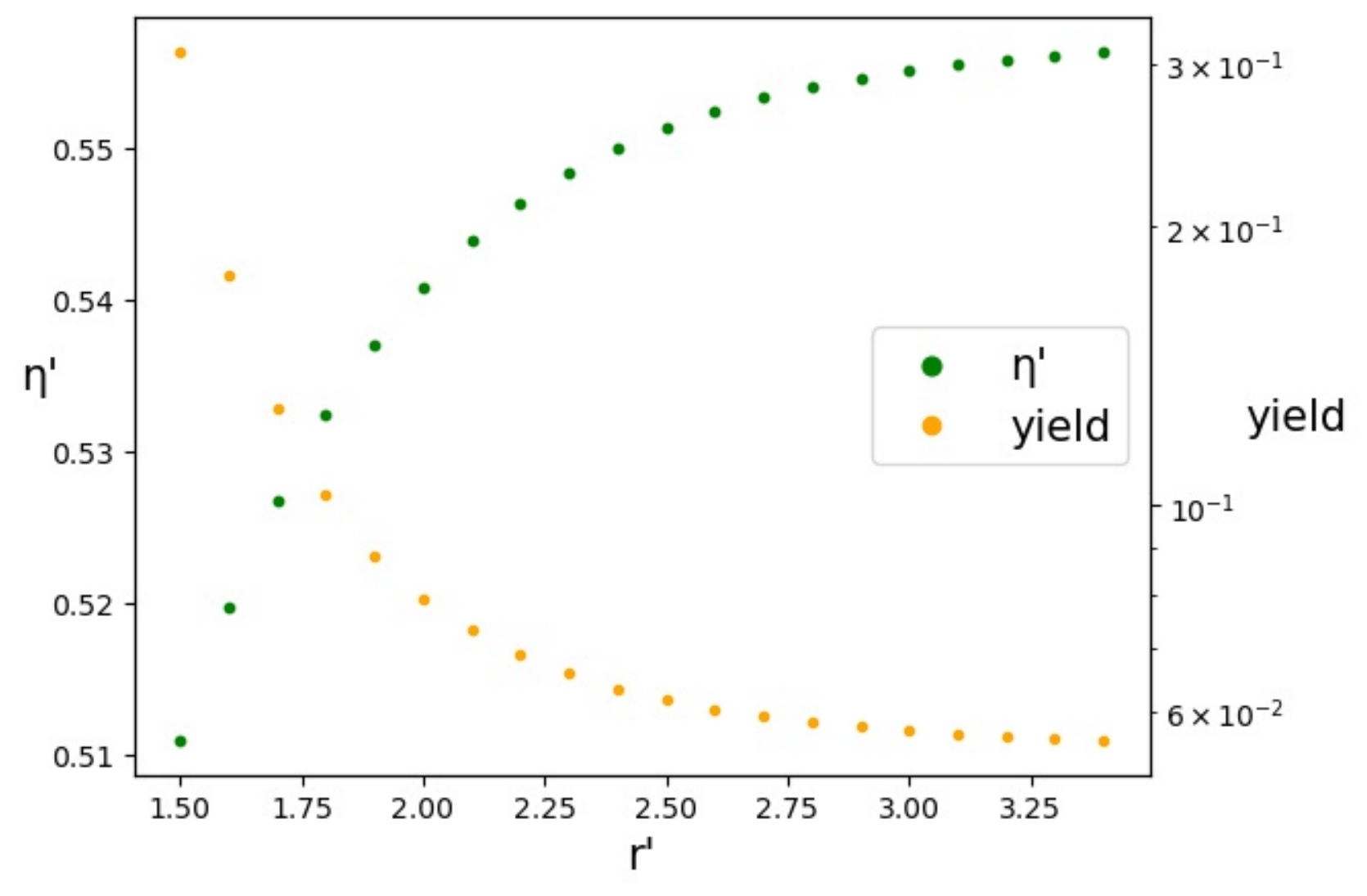} 
		}
		\DeclareGraphicsExtensions.
		\caption{\label{fig:3} Performance of our mitigation method with typical parameters. $r$ is input SMSSs' squeezing strength. $r^{\prime}$ is the squeezing strength to be simulated.  $K$ is the number of input SMSSs. $N_{\text{0}}$ is cut-off photon number. $\eta $ is the overall transmission rate. We use the results of GBS devices with SMSSs squeezing strength $r$ to simulate that of $r'$ through the post-selection method of protocol 1. Red points correspond to $\eta'$ which is the effective transmission rate in final samples. Orange points correspond to yield which is the percentage of preserved outputs.}
	\end{figure}

	In the following paragraphs, we evaluate the effect of the post-selection method by numerical simulations using the parameters of two recent GBS experiments~\cite{madsen_quantum_2022,zhong_phase-programmable_2021}. As we stated above, these numerical simulations are performed based on the analytical formulas in subsection~\ref{sec:41}. The parameters in Fig \ref{fig:31} correspond to the GBS experiment with the highest total photon number measured so far~\cite{madsen_quantum_2022}, in which the squeezing strength of the input squeezed state is $r=1.1$, the number of input squeezed states is $216$, the overall transmission rate of the device is about $ 0.32$, and the maximum photon number of the measured sample is $219$. We take the truncated photon number as $219$. As can be seen from figure \ref{fig:31}, after post-selection, the equivalent transmission rate $\eta'$ of the sample increases with the increase in the squeezing strength $r'$ of the simulated target input squeezed states. When $r'=2.6$, the equivalent transmission rate becomes $\eta'\approx 0.45$, which is about $50\%$ higher than the original transmission rate, while the yield is about $5\times 10^{-7}$. The low yield is mainly due to the low probability of experimentally observing samples with total photon number $N=219$, and should not be simply considered as a defect of the post-selection method. In addition, according to~\cite{madsen_quantum_2022}, the number of samples generated per second in the experiment is about $2\times10^{4}$, which means that if $r'=2.6$ is chosen, then the experimental apparatus can generate a sample in an average of one hundred seconds under the effect of post-selection. If we consider post-selection based on the existing samples, then we can guarantee that $100\%$ of the samples with a total photon number of $N=219$ are retained (which is most advantageous for claiming quantum advantage), while the total number of samples will be reduced to about $5\times10^{-7}$ times the original one.

	The parameters in figure \ref{fig:32} correspond to those in another GBS experiment~\cite{zhong_phase-programmable_2021}. In this GBS experiment, the squeezing strength of the input squeezed state is $r\approx1.4$, the number of input squeezed states is $50$, the overall transmission rate of the device is about $0.5$, and the maximum photon number of the measured sample is $113$. We take the truncated photon number as $113$. As can be seen from Fig \ref{fig:32}, after the post-selection, the equivalent transmission rate $\eta'$ of the sample increases with the increase in the squeezing strength $r'$ of the simulated target input squeezed states. When $r'= 2.5$, the equivalent transmission rate becomes $\eta'\approx 0.55$, which is about $10\%$ higher than the original transmission rate, while the yield is about $0.617$. The reason for the limited improvement in the equivalent transmission rate here is that, as we discussed at the beginning of this subsection, the mitigation of photon loss by the post-selection method decreases if the actual squeezing strength and transmission rate used in the experiment are increased. Although in this case, the post-selection results in a limited improvement of the sample equivalent transmission rate, as we will point out in subsection~\ref{sec:62}, our method can help experimenters increase the circuit depth of their experimental equipment to meet the challenges from classical simulations~\cite{oh_classical_2022}. The high yield reflects the high probability of observing a sample with total photon number $N=113$ experimentally. Of course, the total photon number is not always equal to the total number of detector clicks. We will demonstrate in subsection~\ref{sec:52} that the limited PNR capability does not negatively affect the performance of post-selection method.

	\subsection{comparison with the existing methods}\label{sec:43}
	Two error mitigation method to mitigate the photon loss error of GBS devices (quantum optical devices) which are given in~\cite{su_error_2021}.
	However, they are very different in terms of both effectiveness and application scenarios. The most significant differences are, according to the example given in section 5 of the article~\cite{su_error_2021}, those methods are mainly applicable to the case of small photon loss in GBS devices and become unworkable when the effect of photon loss in GBS devices is severe. In contrast, according to our analysis in subsection~\ref{sec:4}, the post-selection method can be used under conditions with large photon loss.
	
	Moreover, the goal of the methods given in~\cite{su_error_2021} is very different from post-selection method. The methods in the article~\cite{su_error_2021} require measuring the probability distribution of the samples first and then obtaining the probability distribution after error mitigation. The post-selection method, on the other hand, does not require collecting and estimating the probability of the samples. In fact, the post-selection method works directly on the samples and the result of the post-selection process is only the error-mitigated samples. 
	For GBS systems with a large number of modes, it is difficult to precisely measure the probability that a particular sample will occur in a GBS experiment. This is because it requires collecting a huge number of samples. The solution given in the article is to no longer do error mitigation for the probability of occurrence of a specific sample, but for the probability of occurrence of a class of samples (i.e. the coarse grained probability). However, if error mitigation is done only for the probability of generation of a particular class of samples, it remains to be clarified what effect this error mitigation approach has on the samples generated by the GBS experiment as a whole and whether it is significant in enhancing the quantum dominance results of the GBS experiment. However, unlike previous methods, the value of the post-selection method for improving the quantum advantage results of GBS experimental devices is more clear, and we will show this result in section~\ref{sec:6}.

	Another difference between the post-selection method and the methods in~\cite{su_error_2021} is that the methods in~\cite{su_error_2021} work for general input states. The post-selection method does not work for arbitrary quantum states of the input. However, in Appendix~\ref{ap:1} we will show that when the photon number distribution of the input state satisfies the following relation
	\begin{equation}
		\frac{P^{(a)}(\bar{n})}{P^{(b)}(\bar{n})}=\frac{P^{(a)}(\bar{m})}{P^{(b)}(\bar{m})},
	\end{equation}
	the post-selection method is still applicable, where $a,b$ are the parameters describing the input quantum states.

	\section{other experimental imperfections and their impact on the post-selection process}\label{sec:5}
	\subsection{non-uniform loss} \label{sec:51}
	In this subsection we will analyze the effect of post-selection on non-uniform loss processes. Although uniform loss is a good approximation for the current GBS experiments, non-uniform loss will inevitably be presented in the experiments~\cite{zhong_quantum_2020,arrazola_quantum_2021,zhong_phase-programmable_2021,madsen_quantum_2022}. By splitting a general non-uniform loss process into a uniform loss process and another non-uniform loss process, we demonstrate that the post-selection method has an error-mitigating effect on the general loss process. In addition, we note that in most cases, the non-uniform loss in GBS experiments is largely attributed to the different detection efficiencies of photon detectors corresponding to different spatial modes. To solve this problem, we give a modified post-selection protocol to specifically deal with the non-uniform loss due to the different detection efficiencies of photon detectors.
	
	By splitting a general loss process into a uniform loss process and another non-uniform loss process, and setting the overall transmission rate $\eta$ in the post-selection protocol to the transmission rate $\eta_u$ of the uniform loss process, we can show that the post-selection reduces the uniform loss while making the non-uniform loss reduced. Proof of this can be found in Appendix \ref{ap:2}.
	
	A large variation in the detection efficiency in the experiment can also lead to a non-uniform loss process for the whole GBS experiment. We propose an improved protocol to deal with this problem.
	The overall transmission rate in the GBS device except for the detectors is denoted as $\eta$, and the detection efficiency of each photon detector is denoted as $\eta_i^d$, where the subscript $i$ refers to the $i$th photon detector.
	The whole loss process can be described by a unitary operator $\hat{U}_{d}$ which includes $M$ actual modes and $M$ loss modes.
	The input-output relationship corresponding to the loss in the whole GBS process is
	\begin{equation}
		\hat{U}_{d} \boldsymbol{a}^{\dagger} \hat{U}_{d}^{\dagger} = \Lambda_{d} \boldsymbol{a}^{\dagger},
	\end{equation}
	where 
	\begin{equation}
		\Lambda_d=\left(
		\begin{matrix} 
			\oplus_{i=1}^M\sqrt{\eta_i}&\oplus_{i=1}^M\sqrt{1-\eta_i}\\
			-\oplus_{i=1}^M\sqrt{1-\eta_i}&\oplus_{i=1}^M\sqrt{\eta_i}\\
		\end{matrix}
		\right),
	\end{equation}
	and $\eta_i = \eta \eta^d_i$.
	Similar to the analysis in Appendix~\ref{ap:1}, the probability distribution of the output sample is
	\begin{small}
		\begin{equation}
			\begin{aligned}
				P_{\text{out}}(\bar{m})&=\sum_{\bar{n},\bar{l}}v(\bar{m}|\bar{n})v^*(\bar{m}|\bar{l})\sum_{\bar{i}\ge\bar{n}}\sum_{\bar{j}\ge\bar{l}}\prod_{p=1}^{M}\prod_{q=1}^{M}\sqrt{P_{\text{in}}(\bar{i})P_{\text{in}}(\bar{j})}\\
				&\quad\sqrt{\binom{i_p}{n_p}(\eta_p)^{n_p}(1-\eta_p)^{i_p-n_p}} 
				\sqrt{\binom{j_q}{l_q}(\eta_q)^{l_q}(1-\eta_q)^{j_q-l_q}} \\
			\end{aligned}.
		\end{equation}
	\end{small}From this equation above, we can see that the protocol \ref{pro:1} still works, as long as we change the post-selection probability to
	\begin{equation}
		P_{\text{post2}}(\vert \bar{n} \vert) = c^{\alpha-| \bar{n}|} \prod_{i=1}^{M}\left(\frac{\eta'_i}{\eta_i}\right)^{n_i},
	\end{equation}
	where  $\eta'_i =1- \left(1-\eta_i\right) \frac{\tanh r}{\tanh r'}$.
	
	\subsection{PNR capability and dark counts}\label{sec:52}
	In this subsection, we will discuss the impact of the detector's limited PNR capability on the post-selection process and how the dark count rate of the detector changes under post-selection. We will show that the post-selection process reduces the sampling error caused by the limited PNR capability of the detectors, but the dark count rate is slightly increased after post-selection.
	
	We assumed in our previous analysis that the total photon number of detected samples can be accurately resolved in GBS experiments, which is not always hold in real GBS experiments. Nevertheless, we will show that the limited PNR capability does not impose a limitation on the use of the post-selection method. Specifically, the post-selection method reduces the number of samples with the wrong photon number in the sample due to the limited PNR capability of the detector. 
	These events that exceed the PNR capability of the detector should all be classified as errors~\cite{quesada_gaussian_2018}.
	Denote the real output pattern in a GBS experiment as $\left|\bar{n}\right\rangle$, while the measured pattern is $\left|\bar{m}\right\rangle$ due to the limited PNR capability of the detector. This error event should have been retained with a probability of $P_{\text{post}}(|\bar{n}|)$ during the post-selection process. But since the total photon number of the detected sample is $|\bar{m}|$ , the actual probability that this error event is retained is $P_{\text{post}}(|\bar{m}|)$. Thus the probability of this error event occurring after post-selection is equivalent to $\frac{P_{\text{post}}\left(|\bar{m}|\right)}{P_{\text{post}}\left(|\bar{n}|\right)}$ times the original probability. Combining with our analysis in subsection \ref{sec:41}, the smaller the total photon number of the sample, the smaller the probability of being retained, so $\frac{P_{\text{post}}\left(\bar{m}\right)}{P_{\text{post}}\left(\bar{n}\right)}<1$. That is, after the post selection process, the number of events exceeding the PNR capability of the detector is reduced. Since the sampling results with quantum advantage in the GBS experiments are composed of the sampling results that are correctly resolved by the photon number detector for the total photon number~\cite{quesada_gaussian_2018}, the post-selection method is thus useful for reducing the errors in the GBS experiments even when the photon detector does not resolve the total photon number of the samples well.
	
	Another detector-related issue in GBS experiments is the inherent dark counts that exists in the detector. The dark counts in the detectors lead to a larger total photon number in the measured sample. This leads to an increase in the probability of that sample being retained by the post-selection process, which in turn equals to an increase in the dark count rate in the experiment. In what follows, we take the threshold detector as our model to calculate the effect of the post-selection process on dark count rate. The dark count rate of a detector is denoted as $p_D$. Whether or not a dark count is generated by a detector is independent of each other. If dark counts occur in a detector, then the probability that the sample is retained is $P_{\text{post}}(N+1)$, where $N$ is the actual total photon number of the sample. 
	This corresponds to increasing the ratio of the occurrence of dark counts to $P_D \left(\frac{P_{\text{post}}(N+1)}{P_{\text{post}}(N)}\right)=p_D\left(\frac{1/c-\left(1 - \eta\right)}{\eta}\right)<p_D\left(1+ \frac{1/\tanh r-1}{\eta}\right)$. And the ratio of dark counts not occurring in this detector is still $1-p_D$. Thus, the equivalent dark count rate becomes
	\begin{align}
		P_D'
		&=\frac{p_DP_{\text{post}}(N+1)}{p_DP_{\text{post}}(N+1)+\left(1-p_D\right) P_{\text{post}}(N)}\\
		&=\frac{p_D\left(\frac{1/c-\left(1-\eta\right)}{\eta}\right)}{p_D\left(\frac{1/c-\left(1-\eta\right)}{\eta}\right)+1-p_D}\\
		&<\frac{p_D\left(1+\frac{1/\tanh r-1}{\eta}\right)}{p_D\left(1+\frac{1/\tanh r-1}{\eta}\right)+1-p_D}.
	\end{align}
	It can be seen that the dark count rate is amplified after a post-selection. However, the amplification of the dark counts is generally not very large, and the dark count rates in the experiments are generally low (about $10^{-3}$ to $10^{-4}$). Thus this generally does not have a significant impact on the experimental results. As we will show in the example of \ref{sec:51}, in many cases the effect of the increase in dark count on the quantum advantage results in GBS experiments is negligible compared to the effect of the decrease in photon number loss.
	
	\section{Further Analysis on Advantage of Post Selection}\label{sec:6}
	\subsection{Advantage in criteria of “non-classicality test”}\label{sec:61}
	In this subsection, we give reasons why we say the post-selection method can enhance the quantum advantage results in GBS experiments. Indeed, recent GBS experiments achieve great success,
	in the sense that those experiments achieve a relatively large number of optical modes
	and observe samples with large total photon numbers (about $100$ to $200$)~\cite{zhong_quantum_2020,zhong_phase-programmable_2021,madsen_quantum_2022}. 
	However, imperfections in the experiments have brought these quantum advantage results into question~\cite{qi_regimes_2020,villalonga_efficient_2022, martinez-cifuentes_classical_2022,oh_classical_2022,shi_effect_2022}.
	In fact, it is not reasonable to claim quantum advantage simply by the maximum number of photons detected in a sample if the probability distribution of the sampling results in the experiment deviates significantly from the theoretical value. According to known results, e.g.~\cite{qi_regimes_2020,oh_classical_2022},
	classical methods can simulate the sampling results when the experimental error exceeds a certain range. These classical methods compare the deviations of simulating the ideal GBS sampling process with classical algorithms and the deviations of the experimentally obtained samples from the ideal sampling results. In general, these deviations can be measured by total variance distance, K-order correlation, maximum likelihood estimation, etc. Various classical approaches to simulate GBS sampling processes using imperfections in experiments have been given, such as~\cite{qi_regimes_2020,villalonga_efficient_2022,martinez-cifuentes_classical_2022,oh_classical_2022,shi_effect_2022}. Therefore, we believe that it is worthwhile to discard a portion of the sample in order to obtain more reliable experimental results to meet the challenges from classical simulations. In the following, we provide some examples as support for our view.
	
	First, we will show that our post-selection approach can make a GBS experiment that could have been effectively approximately simulated by a classical algorithm become incapable of being simulated by that classical algorithm. This so called "non-classicality test"~\cite{qi_regimes_2020} for photonic devices we consider in this example is currently used by a variety of GBS experiments to test whether their experiments achieve a quantum advantage~\cite{qi_regimes_2020,arrazola_quantum_2021,zhong_phase-programmable_2021,madsen_quantum_2022}. The classical algorithm used in this example is presented in the article~\cite{qi_regimes_2020}, where it is shown that a classical algorithm can efficiently simulate the sampling process of a GBS experiment up to an error $\varepsilon$ when the following condition is satisfied:
	\begin{equation}
		\operatorname{sech}\left\{\frac{1}{2} \Theta\left[\ln \left(\frac{1-2 q_D}{\eta e^{-2r}+1-\eta}\right)\right]\right\}>\mathrm{e}^{-\varepsilon^2/4K},
		\label{eq:43}
	\end{equation} 
	where $\Theta$ is the ramp function $\Theta\left(x\right) =\mathrm{max}\left(x,0\right)$, $r$ is the squeezing strength of the input squeezed states, $\eta$ is the overall transmission rate (from the optical sources to the detectors), $K$ is the total number of input squeezed states and $q_D$ is the dark count rate of the photon detectors. When the inequality above has no solution for $\varepsilon\in\left[0,1\right]$, it means that the GBS experiment pass "the non-classicality test". Denote $\varepsilon_0$ as the minimal value which satisfies the inequality given in Eq.~(\ref{eq:43}), that is:
	\begin{align}
		\operatorname{sech}\left\{\frac{1}{2} \Theta\left[\ln \left(\frac{1-2 q_D}{\eta e^{-2 r}+1-\eta}\right)\right]\right\}=e^{-\varepsilon_0^2 / 4 K}.
	\end{align}
	According to \cite{qi_regimes_2020}, when $\varepsilon_0$ is larger than 1, this classical simulation algorithm fails.
	
\begin{figure}[htbp]
		\centering
		\subfigure[]{\label{fig:41}
			\includegraphics[width=0.95\linewidth]{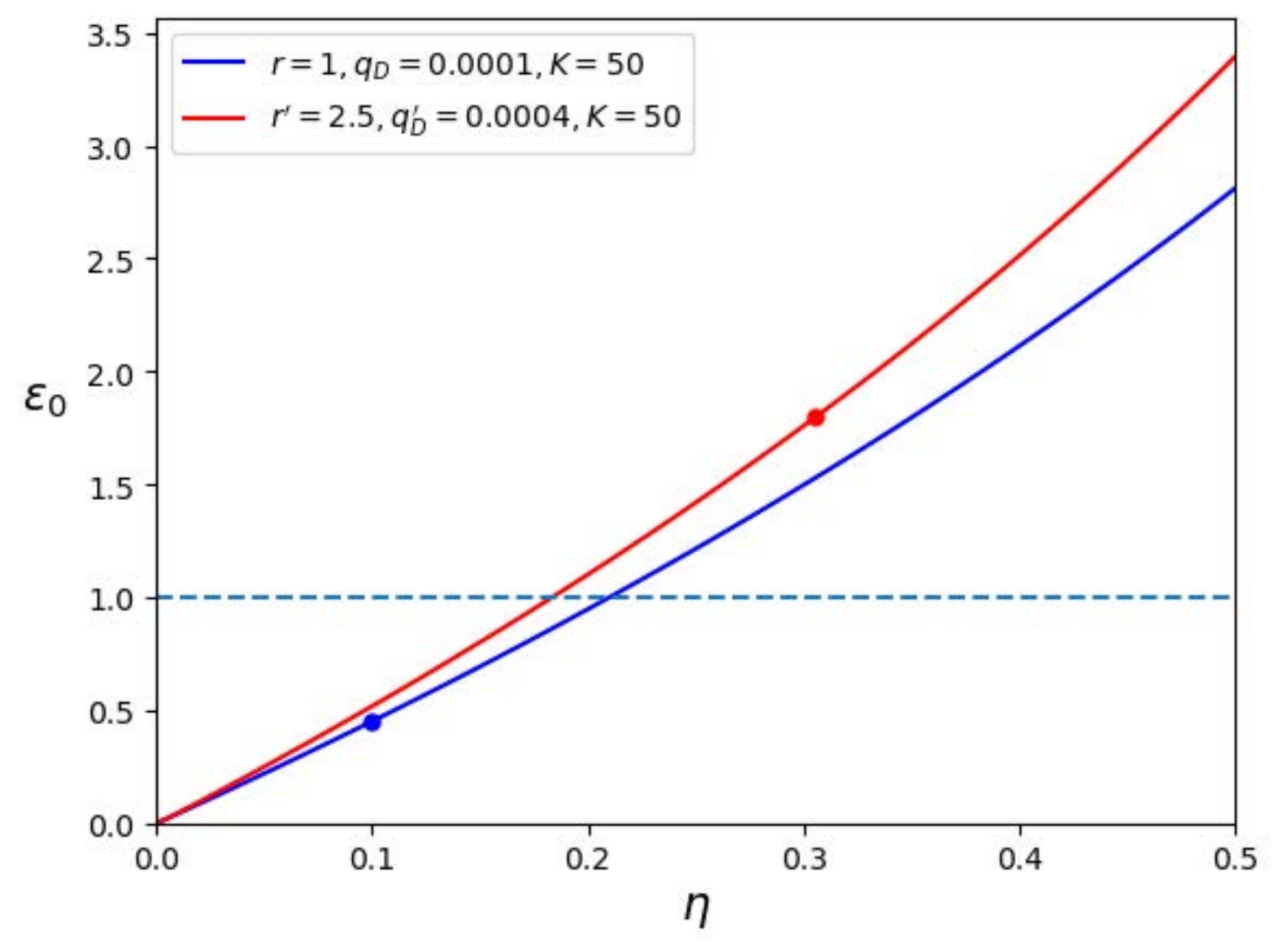} 
		}
		\subfigure[]{\label{fig:42}
			\includegraphics[width=0.95\linewidth]{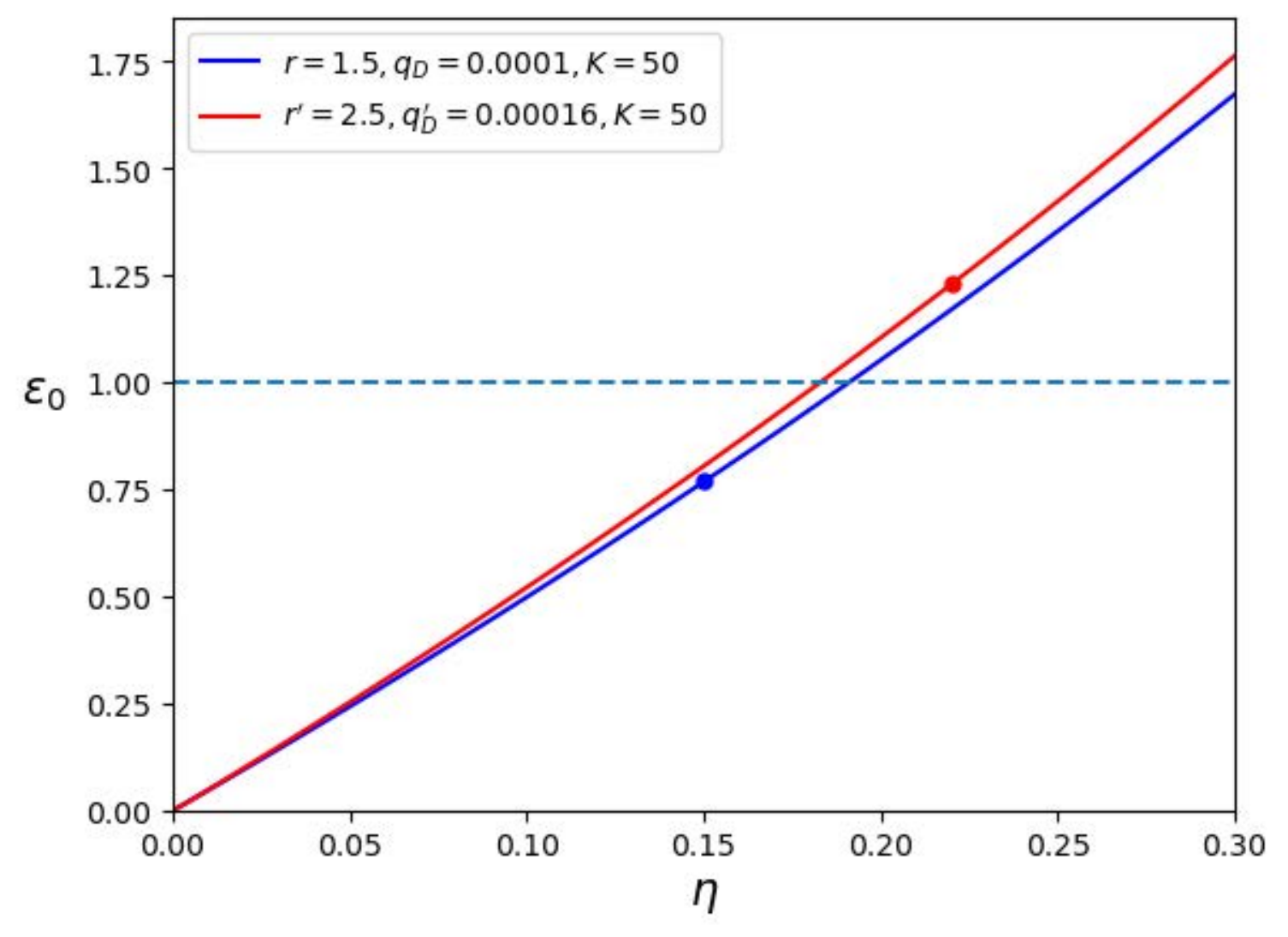} 
		}
		\DeclareGraphicsExtensions.
		\caption{\label{fig:4} Simulation errors of the classical algorithm given in~\cite{qi_regimes_2020} for different experimental parameters. The horizontal axis corresponds to the transmission rate $\eta$ of the sampling process. The vertical axis corresponds to the error upper bound $\varepsilon_0$ of the classical simulation algorithm. When the error upper bound $\varepsilon_0$ on the vertical axis exceeds $1$ (dashed line in the figure), the classical simulation algorithm fails. The blue curve corresponds to the simulation error of the classical algorithm at different transmission rates for a GBS sampling process. The red curve corresponds to the simulation error of the classical algorithm for the GBS sampling process (corresponding to the blue curve in each figure) after post-selection (the squeezing strength of the simulated state is $r'$) at different transmission rates. The blue points correspond to the error upper bound $\varepsilon_0$ of the  classical simulation algorithm for the GBS experiment at the transmission rate $\eta$. The red points correspond to the error upper bound $\varepsilon_0'$ of the classical simulation algorithm for the GBS experiments (corresponding to the blue points in each figure) after post-selection (the squeezing strength of the simulated state is $r'$). (a). $r=1$, $K=50$, $q_D=0.0001$, $r'=2.5$. (b). $r=1.5$, $K=50$, $q_D=0.0001$, $r'=2.5$.}
		
	\end{figure}

	In Fig~\ref{fig:4}, we give two examples to show that the post-selection method can make a GBS sampling process that could be approximately simulated by a classical algorithm become unable to be simulated by the same classical algorithm.  The blue line shows the error of the classical simulation of a GBS experiment under different overall transmission rates $\eta$. The squeezing strength of the input squeezed states in this GBS experiment is $r$, the dark count rate of the detectors is $q_D$, and the number of input squeezed states is $K$. The red line shows the error of the classical simulation of the whole GBS sampling process (including post-selection process) under different equivalent overall transmission rates $\eta'$. The squeezing strength of the simulated target squeezed states for the post-selection process is set to $r'$. The vertical axis is the upper bound of the simulation error $\varepsilon_0$ of the classical simulation algorithm. The horizontal axis corresponds to the overall transmission rate $\eta$. Here we use the formula in subsection~\ref{sec:52} to calculate the change in the dark count rate $q_D$ occurring in the GBS sampling process after post-selection. The equivalent dark count rate after the post selection process is denoted as $q_D'$.  In case corresponds to Fig~\ref{fig:41}, where $r=1$, $K=50$, $q_D=0.0001$, $r'=2.5$, when the overall transmission rate is $\eta=0.1$, the sampling process can be simulated by the classical algorithm with an error of at most $45\%$ (corresponding to the blue dots in the figure). If the post-selection is done for this sampling process, this GBS sampling process will no longer be able to be simulated by this classical algorithm (corresponding to the red dots in the figure), as $\varepsilon_0$ exceeds $1$. Similarly, in case corresponds to Fig~\ref{fig:42}, where $r=1.5$, $K=50$, $q_D=0.0001$, $r'=2.5$, when the overall transmission rate $\eta=0.15$, this sampling process can be simulated by the classical algorithm with an error of at most $77\%$ (corresponding to the blue dots in the figure). If the post-selection is done for this sampling process, this GBS sampling process will no longer be able to be simulated by this classical algorithm (corresponding to the red dots in the figure),  as $\varepsilon_0$ exceeds $1$.

	These examples show that the post-selection method can help to improve the robustness of current GBS devices. We need to make it clear that our post-selection method is not applicable to the classical simulation algorithm in the examples above, because the errors in this algorithm do not fit the model of photon loss. Of course, there is no way to know whether one will discover new algorithms to enhance the simulation power of classical computers. However, whatever further improvements may be made to the classical algorithm, our analysis and examples have shown the value of using post-selection method to reduce photon loss for enhancing the quantum advantage results of GBS devices.
	
	\subsection{Advantage in increasing the GBS circuit depth}\label{sec:62}
	
	One can use the post-selection method to increase the circuit depth of the GBS devices. Let's start by explaining why it is important to increase the circuit depth of GBS devices. One of the problems that GBS experiments face nowadays is that the circuit depth is too shallow~\cite{zhong_quantum_2020,zhong_phase-programmable_2021}. A shallow circuit depth leads to weak correlation between different modes in the GBS experimental results. This is exploited in a recent article~\cite{oh_classical_2022}, where a classical algorithm is designed to simulate GBS experiments with shallow circuit depth. This article gives an example where, due to the shallow circuit depth, a classical simulation method (with sub-exponential time complexity) yields samples with a lager likelihood of sampling compared to the probability distribution of the ideal sampling results than those obtained in the GBS experiment~\cite{zhong_phase-programmable_2021}. However, with the current technology, a major difficulty in increasing the circuit depth is that the transmission rate decreases exponentially with the number of layers in the circuit. If the transmission rate of one layer in the circuit is $\eta_0$, then the transmission rate of $D$ layer is $\eta=\left(\eta_0\right)^D$. This phenomenon of exponential decay of the transmission rate limits the circuit depth of the GBS experiments nowadays. Post-selection method can help people mitigate the decrease in transmission rate and increase the circuit depth of GBS devices.

	Use the experimental parameters in~\cite{zhong_quantum_2020} to make an estimate. If a GBS experiment with an overall transmission rate $\eta_1=0.5$ and a circuit depth of $D$, then the transmission rate for each layer is about $\eta_0=(\eta)^{1/D}$. Increasing the circuit depth to $2D$ layers, the corresponding overall transmission rate is $\eta_2=(\eta_0)^{2D}=0.25$. If the squeezing strength of the input squeezed states is $r=1$ and the squeezing strength of the post-selection simulated target squeezed states is $r'=2.5$, then the equivalent transmission rate of the GBS sampling process after post-selection is $\eta'=0.42$, which is not much lower than the original overall transmission rate, and the circuit depth is increased to twice the original depth. This is beneficial to enhance the quantum advantage results in the current GBS experiment. Because it may help the samples obtained from the GBS experiment to get better performance in likelihood test than the classical simulation results~\cite{oh_classical_2022}. 
	
	This example again shows the role of the post-selection method in enhancing the quantum advantage results in the present GBS experiment. If the experiment is accompanied by a reduction in the squeezing strength of the input squeezed states, the experimenter can increase the circuit depth of the GBS device even more. Of course, in order to maintain a large total photon number for the samples obtained from the GBS experiment, the number of input squeezed states needs to be increased. However, overall, these efforts are worthwhile for enhancing the quantum advantage of GBS experiments.

	\section{Summary}\label{sec:7}
	In summary, we have presented a method to suppress the effect of photon loss in GBS devices. Our method takes post-selection of the results of real experiments so as to obtain high-quality sampling results of another source.  The error mitigation method proposed in this paper is easy to be performed in nowadays’GBS devices, as no hardware modification is needed.  Further, theoretical analysis shows that this post-selection method is robust against photon loss. 
	
	The effects of the post selection process on some other typical types of errors errors in GBS experiments  are also carefully analyzed, including non-uniform loss, the detector's limited PNR capability, and dark counting. We demonstrate that the errors due to both the non-uniform loss and the limited PNR capability of the detector are reduced by the post selection process, while the dark count rate is slightly increased by the post-selection process. Since the experimental dark count rate is generally small, the increase of dark count rate has less impact on the GBS experimental results compared to the increase in the total transmission rate. This statement is supported with the examples in subsection~\ref{sec:61}.
	
	This post selection method can improve GBS devices’ performance and enhence the quantum advantage results. We demonstrate this with some examples. We show that the post-selection approach can make GBS experiments that previously failed the non-classical test to pass the non-classical test. This means that GBS experiments that were originally judged to have no quantum advantage can become potentially quantum advantageous by post-selection of samples. In addition, we show that our method can help improve the circuit depth of GBS devices, which is also important for enhancing the quantum advantage results of GBS experiments. Thus, we believe that the post-selection method is of valuable for enhancing the quantum advantage results of existing GBS devices, as well as for the further development of the GBS-based quantum algorithms.
	
	\begin{acknowledgements} 
		We thank Ji-Qian Qin, Yun-Long Yu, Fan Yang and Zong-Wen Yu for valuable discussions.
		We \mbox{acknowledge} the financial support in part by National Natural Science Foundation of China grant No.11974204 and No.12174215.
	\end{acknowledgements}
	
	\appendix

	\section{different input sources}\label{ap:1}
	Similar to the analysis in theorem 2, if the input quantum state in a passive linear optical network is measured under the Fock basis, the probability distribution is $P^{a}(n)=\mathcal{N}\sum_{n}f(a)^{n}$, where $\mathcal{N}$ is the normalization factor and $a$ corresponds to the parameter describing that input quantum state. 
	If several input ports are input with this type of quantum states, then the probability distribution of the input quantum state in the Fock basis is $P^{\{\bar{a}\}}(\bar{n})=\mathcal{N}'\sum_{\bar{n}}f(\bar{a})^{\bar{n}}$, where $\mathcal{N} '$ is the normalization factor and $\bar{a}=a_1\dots a_M$ corresponds to the parameters describing the $M$-mode quantum state of the input.
	Assuming that the overall transmission rate of the passive linear optical network is $\eta$, the probability of getting a sample $\bar{n}$ detected at the output in the Fock basis is
	\begin{equation}
		P_{\text{out}}^{\left\{\bar{a},\eta\right\}}\left(\bar{n}\right)=
		\sum_{\bar{m},\bar{l}}v_1(\bar{n}\mid\bar{m})v_1^*(\bar{n}\mid\bar{l})
		\sqrt{P_{\text{in}}^{\left\{\bar{a},\eta\right\}}(\bar{m})P_{\text{in}}^{\left\{\bar{a},\eta\right\}}(\bar{l})},
	\end{equation}
	where
	\begin{equation}
		P^{\{\bar{a},\eta\}}_{\text{in}}(\bar{m})=\sum_{\bar{i}\ge\bar{m}}\prod_{p=1}^{M}
		\binom{i_p}{m_p}\eta^{m_p}(1-\eta)^{i_p-m_p}P^{\{\bar{a}\}}(\bar{i}).
	\end{equation}
	If there is
	\begin{equation}
		\frac{P^{(a)}(\bar{n})}{P^{(b)}(\bar{n})}=\frac{P^{(a)}(\bar{m})}{P^{(b)}(\bar{m})},
		\label{eq:B3}
	\end{equation}
	where $|\bar{m}|=|\bar{n}|$, then
	\begin{equation}
		\begin{aligned}
			&P^{\{\bar{a},\eta\}}_{\text{in}}(\bar{m}) \frac{P^{\{\bar{b}\}}(|\bar{n}|)}{P^{\{\bar{a}\}}(|\bar{n}|)}\left(\frac{\eta'}{\eta}\right)^{|\bar{n}|}\\\
			&=\sum_{\bar{i}\ge\bar{m}} \prod_{p=1}^{M}\binom{i_p}{m_p}\left(\eta'\right)^{m_p}(1-\eta')^{i_p-m_p}P^{\{\bar{b}\}}(\bar{i })\\
			&=P^{\{\bar{b},\eta'\}}_{\text{in}}(\bar{m}).
		\end{aligned}
	\end{equation}
	This proves that if the probability distribution of the input state in the Fock basis satisfies the Eq.~(\ref{eq:B3}), then performing post-selection with probability
	\begin{equation}
		P(|\bar{n}|)=\mathcal{N}''\frac{P^{\{{\bar{b}\}}(|\bar{n}|)}}{P^{\{\bar{a}\}}(|\bar{n}|)}\left( \frac{\eta'}{\eta}\right)^{|\bar{n}|}
	\end{equation}
	can reduce the effect of photon loss on the sampling process, where $\mathcal{N}''$ makes $P(N_0)=1$ and $N_0$ is the cut-off photon number corresponding to the post-selection.
	
	In the following we give an example where the input quantum state matches (\ref{eq:B3}). 
	The article~\cite{rohde_evidence_2015} argues that if the cat states are input into a passive linear optical network and measured at the output ports with photon number detectors, then this sampling process may also be classically difficult to simulate. The cat state is defined as
	\begin{equation}
		\left|\psi\right>=\frac{\left|\alpha\right>+\left|-\alpha\right>}{\sqrt{2+2\operatorname{e}^{-2 \left|\alpha\right|^2}}},
	\end{equation}
	where $\left|\alpha\right>$ is the coherent state.
	The photon number distribution of the cat state is
	\begin{equation}
		\begin{aligned}
			P^{(\alpha)}(n)=&\left|\left<n|\psi\right>\right|^2=
			\frac{1}{2+2\operatorname{e}^{-2\left|\alpha\right|^2}}\\
			&\times\left(2\operatorname{e}^{-\left|\alpha\right|^2}\frac{\left|\alpha\right|^{2n}}{n!}+
			2\operatorname{e}^{-\left|\alpha\right|^2}\frac{\left(-1\right)^n\left|\alpha\right|^{2n}}{n!}
			\right).
		\end{aligned}
	\end{equation}
	The calculation gives
	\begin{equation}
		\frac{P^{(\alpha)}(n)}{P^{(\beta)}(n)}=\frac{1+\operatorname{e}^{-2\left|\alpha\right|^2}}{1+\operatorname{e}^{-2\left|\alpha\right|^2}}\frac{\operatorname{e}^{-\left|\alpha\right|^2}\left|\alpha\right|^{2n}}{\operatorname{e}^{-\left|\beta\right|^2}\left|\beta\right|^{2n}}
	\end{equation}
	It can be seen that the Eq.~(\ref{eq:B3}) is satisfied.
	
	\section{non-uniform loss}\label{ap:2}
	Consider an $M$-mode interferometer (passive linear optical network), if the interferometer is lossy, then the input-output relation is~\cite{brod_classical_2020,garcia-patron_simulating_2019,oszmaniec_classical_2018}
	\begin{align}
		\hat{b}=A\hat{a}+\sqrt{I-AA^{\dagger}}\hat{e},
	\end{align}
	where $\hat{e}$ represents the environmental modes which will be traced out, and $A$ is a complex matrix satisfying $AA^{\dagger}\le I$. The matrix $A$ can be decomposed (singular decomposition) as $A = V_1 D V_2$, where $V_1$and $V_2$ are unitary matrices, and $D = \mathrm{diag}\{\sqrt{\eta_1},\sqrt{\eta_2},...,\sqrt{\eta_M}\}$ with $\eta_i \in [0,1]$.

	Denote $\eta_{u}=$ $\max \left(\eta_{i}\right)$ for $i=1,2, \ldots, M$.  A uniform loss layer with transmission rate $\eta_{u}$ can be extracted from the lossy passive linear optical network. This can be seen from:
	\begin{align}
		A=\sqrt{\eta}_{u} I\left(\frac{1}{\sqrt{\eta_{u}}}\right) V_1 D V_2 .
	\end{align}
	Thus the whole physical process is equivalent to that the input SMSSs first go through a passive linear optical network with a nonuniform loss described by $A_{1}=$ $\left(\frac{1}{\sqrt{\eta_{u}}}\right) V_1 D V_2$ and then a uniform loss channel described by $A_{2}=\sqrt{\eta}_{u} I$.
	
	The probability that an output pattern $\bar{m}$ is obtained is 
	\begin{equation}
		\begin{aligned}
			P_{\text{out}}(\bar{m}) =&\left\langle\bar{m}\right|\mathrm{Tr}_{2M}\{\hat{U}_1\hat{U}_u\hat{U}_n\hat{U}_2\left(\hat{\rho}_{\text{in}}\otimes\left|\bar{0}\left\rangle\right\langle\bar{0}\right|\right)\\
			&\times\hat{U}_2^\dagger\hat{U}_n^\dagger\hat{U}_u^\dagger\hat{U}_1^\dagger\}\left|\bar{m}\right\rangle.
		\end{aligned}
	\end{equation}
	Since a general loss process is split into a uniform loss process and a non-uniform loss process, there are $2M$ loss modes.
	After simplification, we get
	\begin{equation}
		\begin{aligned}
			P_{\text{out}}(\bar{m}) =&\sum_{\bar{n},\bar{l},\bar{k},\bar{d},\bar{z},\bar{g},\bar{b},\bar{i},\bar{j}}
			v_1(\bar{m}\mid\bar{n})v_1^*(\bar{m}\mid\bar{l})v_2(\bar{k}\mid\bar{d})\\
			&\times v_2^*(\bar{z}\mid\bar{g})\sqrt{P^{\{r_i\}}_{\text{in}}(\bar{d})P^{\{r_i\}}_{\text{in}}(\bar{g})}\left\langle \bar{n} , \bar{b}\left|\hat{U}_u \right|\bar{i},\bar{0} \right\rangle \\
			&\times \left\langle \bar{j} , \bar{0}\left|\hat{U}_u^\dagger \right|\bar{l},\bar{p}\right\rangle
			\left\langle \bar{i} , \bar{c}\left|\hat{U}_n \right|\bar{k},\bar{0} \right\rangle
			\left\langle \bar{z} , \bar{0}\left|\hat{U}_n^\dagger \right|\bar{j},\bar{v} \right\rangle
			,
		\end{aligned}
	\end{equation}
	where $v_2(\bar{k}|\bar{d})=\langle\bar{k}|\hat{U}_2|\bar{d}\rangle$ and $v^*_2(\bar{z}\mid\bar{g})=\langle\bar{g}|\hat{U}_2^\dagger|\bar{z}\rangle$.
	Since the passive linear optical network keeps the total photon number in the input and output states unchanged, we have$|\bar{m}|=|\bar{n}|=|\bar{l}|$, $|\bar{k}|=|\bar{d}|$, $|\bar{z}|=|\bar{g}|$, $|\bar{k}|=|\bar{c}|+|\bar{i}|$, $|\bar{z}|=|\bar{j}|+|\bar{v}|$,  $|\bar{j}|=|\bar{p}|+|\bar{l}|$, $|\bar{i}|=|\bar{n}|+|\bar{b}|$.

	Next, take the transmission rate $\eta_u$ of the uniform loss layer as the overall transmission rate to do the post-selection in protocol 1, and the probability distribution of the samples obtained by post-selection in the case of non-uniform loss is (without considering cut-off)
	
	\begin{equation}
		\begin{aligned}
			P_{\text{out}}(\bar{m})P_{\text{post}}(\bar{m})
			=&\mathcal{N}\sum_{\bar{n},\bar{l}} v_1(\bar{m}\mid\bar{n})v_1^*(\bar{m}\mid\bar{l})v_2(\bar{k}\mid\bar{d})\\
			&\times v_2^*(\bar{z}\mid\bar{g})\\ 
			&\sum_{\bar{i} \geq \bar{n}} \sum_{\bar{j} \geq \bar{l}}
			\sum_{\bar{v},\bar{k}\ge\bar{i},\bar{z}\ge\bar{j}}
			\sqrt{P_{\text{in}}^{\{r_i'\}}(\bar{k})P_{\text{in}}^{\{r_i'\}}(\bar{z})}		\\
			&\times\sqrt{
				\binom{\bar{k}}{\bar{i}}
				\left( \sqrt{\eta_p} \right)^{\bar{i}}
				\left( \sqrt{c(1-\eta_p)} \right)^{\bar{k}-\bar{i}}
			}\\	
			&\times
			\sqrt{
				\binom{\bar{z}}{\bar{j}}
				\left( \sqrt{\eta_p} \right)^{\bar{j}}
				\left( \sqrt{c(1-\eta_p)} \right)^{\bar{z}-\bar{j}}
			}\\
			&\times\sqrt{\left(\begin{array}{c}
					\bar{i} \\
					\bar{n}
				\end{array}\right)(\eta_{u}')^{\bar{n}}\left(1-\eta_{u}'\right)^{\bar{i}-\bar{n}}}
			\\
			&\times\sqrt{\left(\begin{array}{c}
					\bar{j} \\
					\bar{l}
				\end{array}\right) (\eta_{u}')^{\bar{l}}\left(1-\eta_{u}'\right)^{\bar{j}-\bar{l}}},
		\end{aligned}
	\end{equation}
	where $\mathcal{N}$ is the normalization factor.
	From this equation above, it can be seen that post-selection reduces the uniform loss while making the non-uniform loss have less effect on the sampling results.
	As, before the post-selection process, the non-uniform loss layer reduces the state $|\bar{k}\rangle$ to $|\bar{i}\rangle$ with the probability
	\begin{equation}
		\mathrm{P}_{n}(\bar{k},\bar{i})=\binom{\bar{k}}{\bar{i}}
		\left( \sqrt{\eta_p} \right)^{\bar{i}}
		\left( \sqrt{1-\eta_p} \right)^{\bar{k}-\bar{i}},
	\end{equation}
	And after the selection, the state $|\bar{k}\rangle$ is reduced to $|\bar{i}\rangle$ with the probability
	\begin{equation}
		\mathrm{P}_{n}'(\bar{k},\bar{i})=\binom{\bar{k}}{\bar{i}}
		\left( \sqrt{\eta_p} \right)^{\bar{i}}
		\left( \sqrt{c(1-\eta_p)} \right)^{\bar{k}-\bar{i}}.
	\end{equation}
	Since $\sqrt{c(1-\eta_p)}<\sqrt{1-\eta_p}$, we find $\mathrm{P}_{n}'(\bar{k},\bar{i})<\mathrm{P}_{n}(\bar{k},\bar{i})$. This corresponds to a reduction in the proportion of the output samples that is affected by non-uniform loss. Therefore, post-selection makes the effect of non-uniform loss on output samples reduced.

	\bibliographystyle{apsrev4-1}
	\bibliography{refs}

\begin{thebibliography}{45}%
\makeatletter
\providecommand \@ifxundefined [1]{%
 \@ifx{#1\undefined}
}%
\providecommand \@ifnum [1]{%
 \ifnum #1\expandafter \@firstoftwo
 \else \expandafter \@secondoftwo
 \fi
}%
\providecommand \@ifx [1]{%
 \ifx #1\expandafter \@firstoftwo
 \else \expandafter \@secondoftwo
 \fi
}%
\providecommand \natexlab [1]{#1}%
\providecommand \enquote  [1]{``#1''}%
\providecommand \bibnamefont  [1]{#1}%
\providecommand \bibfnamefont [1]{#1}%
\providecommand \citenamefont [1]{#1}%
\providecommand \href@noop [0]{\@secondoftwo}%
\providecommand \href [0]{\begingroup \@sanitize@url \@href}%
\providecommand \@href[1]{\@@startlink{#1}\@@href}%
\providecommand \@@href[1]{\endgroup#1\@@endlink}%
\providecommand \@sanitize@url [0]{\catcode `\\12\catcode `\$12\catcode
  `\&12\catcode `\#12\catcode `\^12\catcode `\_12\catcode `\%12\relax}%
\providecommand \@@startlink[1]{}%
\providecommand \@@endlink[0]{}%
\providecommand \url  [0]{\begingroup\@sanitize@url \@url }%
\providecommand \@url [1]{\endgroup\@href {#1}{\urlprefix }}%
\providecommand \urlprefix  [0]{URL }%
\providecommand \Eprint [0]{\href }%
\providecommand \doibase [0]{http://dx.doi.org/}%
\providecommand \selectlanguage [0]{\@gobble}%
\providecommand \bibinfo  [0]{\@secondoftwo}%
\providecommand \bibfield  [0]{\@secondoftwo}%
\providecommand \translation [1]{[#1]}%
\providecommand \BibitemOpen [0]{}%
\providecommand \bibitemStop [0]{}%
\providecommand \bibitemNoStop [0]{.\EOS\space}%
\providecommand \EOS [0]{\spacefactor3000\relax}%
\providecommand \BibitemShut  [1]{\csname bibitem#1\endcsname}%
\let\auto@bib@innerbib\@empty
\bibitem [{\citenamefont {Nielsen}\ and\ \citenamefont
  {Chuang}(2010)}]{nielsen_quantum_2010}%
  \BibitemOpen
  \bibfield  {author} {\bibinfo {author} {\bibfnamefont {M.~A.}\ \bibnamefont
  {Nielsen}}\ and\ \bibinfo {author} {\bibfnamefont {I.~L.}\ \bibnamefont
  {Chuang}},\ }\href@noop {} {\emph {\bibinfo {title} {Quantum computation and
  quantum information}}},\ \bibinfo {edition} {10th}\ ed.\ (\bibinfo
  {publisher} {Cambridge University Press},\ \bibinfo {address} {Cambridge ;
  New York},\ \bibinfo {year} {2010})\BibitemShut {NoStop}%
\bibitem [{\citenamefont {Preskill}(2018)}]{preskill_quantum_2018}%
  \BibitemOpen
  \bibfield  {author} {\bibinfo {author} {\bibfnamefont {J.}~\bibnamefont
  {Preskill}},\ }\href {\doibase 10.22331/q-2018-08-06-79} {\bibfield
  {journal} {\bibinfo  {journal} {Quantum}\ }\textbf {\bibinfo {volume} {2}},\
  \bibinfo {pages} {79} (\bibinfo {year} {2018})}\BibitemShut {NoStop}%
\bibitem [{\citenamefont {Shor}(1994)}]{shor_algorithms_1994}%
  \BibitemOpen
  \bibfield  {author} {\bibinfo {author} {\bibfnamefont {P.}~\bibnamefont
  {Shor}},\ }in\ \href {\doibase 10.1109/SFCS.1994.365700} {\emph {\bibinfo
  {booktitle} {Proceedings 35th {Annual} {Symposium} on {Foundations} of
  {Computer} {Science}}}}\ (\bibinfo {year} {1994})\ pp.\ \bibinfo {pages}
  {124--134}\BibitemShut {NoStop}%
\bibitem [{\citenamefont {Arute}\ \emph {et~al.}(2019)\citenamefont {Arute},
  \citenamefont {Arya}, \citenamefont {Babbush}, \citenamefont {Bacon},
  \citenamefont {Bardin}, \citenamefont {Barends}, \citenamefont {Biswas},
  \citenamefont {Boixo}, \citenamefont {Brandao}, \citenamefont {Buell},
  \citenamefont {Burkett}, \citenamefont {Chen}, \citenamefont {Chen},
  \citenamefont {Chiaro}, \citenamefont {Collins}, \citenamefont {Courtney},
  \citenamefont {Dunsworth}, \citenamefont {Farhi}, \citenamefont {Foxen},
  \citenamefont {Fowler}, \citenamefont {Gidney}, \citenamefont {Giustina},
  \citenamefont {Graff}, \citenamefont {Guerin}, \citenamefont {Habegger},
  \citenamefont {Harrigan}, \citenamefont {Hartmann}, \citenamefont {Ho},
  \citenamefont {Hoffmann}, \citenamefont {Huang}, \citenamefont {Humble},
  \citenamefont {Isakov}, \citenamefont {Jeffrey}, \citenamefont {Jiang},
  \citenamefont {Kafri}, \citenamefont {Kechedzhi}, \citenamefont {Kelly},
  \citenamefont {Klimov}, \citenamefont {Knysh}, \citenamefont {Korotkov},
  \citenamefont {Kostritsa}, \citenamefont {Landhuis}, \citenamefont
  {Lindmark}, \citenamefont {Lucero}, \citenamefont {Lyakh}, \citenamefont
  {Mandrà}, \citenamefont {McClean}, \citenamefont {McEwen}, \citenamefont
  {Megrant}, \citenamefont {Mi}, \citenamefont {Michielsen}, \citenamefont
  {Mohseni}, \citenamefont {Mutus}, \citenamefont {Naaman}, \citenamefont
  {Neeley}, \citenamefont {Neill}, \citenamefont {Niu}, \citenamefont {Ostby},
  \citenamefont {Petukhov}, \citenamefont {Platt}, \citenamefont {Quintana},
  \citenamefont {Rieffel}, \citenamefont {Roushan}, \citenamefont {Rubin},
  \citenamefont {Sank}, \citenamefont {Satzinger}, \citenamefont {Smelyanskiy},
  \citenamefont {Sung}, \citenamefont {Trevithick}, \citenamefont
  {Vainsencher}, \citenamefont {Villalonga}, \citenamefont {White},
  \citenamefont {Yao}, \citenamefont {Yeh}, \citenamefont {Zalcman},
  \citenamefont {Neven},\ and\ \citenamefont {Martinis}}]{arute_quantum_2019}%
  \BibitemOpen
  \bibfield  {author} {\bibinfo {author} {\bibfnamefont {F.}~\bibnamefont
  {Arute}}, \bibinfo {author} {\bibfnamefont {K.}~\bibnamefont {Arya}},
  \bibinfo {author} {\bibfnamefont {R.}~\bibnamefont {Babbush}}, \bibinfo
  {author} {\bibfnamefont {D.}~\bibnamefont {Bacon}}, \bibinfo {author}
  {\bibfnamefont {J.~C.}\ \bibnamefont {Bardin}}, \bibinfo {author}
  {\bibfnamefont {R.}~\bibnamefont {Barends}}, \bibinfo {author} {\bibfnamefont
  {R.}~\bibnamefont {Biswas}}, \bibinfo {author} {\bibfnamefont
  {S.}~\bibnamefont {Boixo}}, \bibinfo {author} {\bibfnamefont {F.~G. S.~L.}\
  \bibnamefont {Brandao}}, \bibinfo {author} {\bibfnamefont {D.~A.}\
  \bibnamefont {Buell}}, \bibinfo {author} {\bibfnamefont {B.}~\bibnamefont
  {Burkett}}, \bibinfo {author} {\bibfnamefont {Y.}~\bibnamefont {Chen}},
  \bibinfo {author} {\bibfnamefont {Z.}~\bibnamefont {Chen}}, \bibinfo {author}
  {\bibfnamefont {B.}~\bibnamefont {Chiaro}}, \bibinfo {author} {\bibfnamefont
  {R.}~\bibnamefont {Collins}}, \bibinfo {author} {\bibfnamefont
  {W.}~\bibnamefont {Courtney}}, \bibinfo {author} {\bibfnamefont
  {A.}~\bibnamefont {Dunsworth}}, \bibinfo {author} {\bibfnamefont
  {E.}~\bibnamefont {Farhi}}, \bibinfo {author} {\bibfnamefont
  {B.}~\bibnamefont {Foxen}}, \bibinfo {author} {\bibfnamefont
  {A.}~\bibnamefont {Fowler}}, \bibinfo {author} {\bibfnamefont
  {C.}~\bibnamefont {Gidney}}, \bibinfo {author} {\bibfnamefont
  {M.}~\bibnamefont {Giustina}}, \bibinfo {author} {\bibfnamefont
  {R.}~\bibnamefont {Graff}}, \bibinfo {author} {\bibfnamefont
  {K.}~\bibnamefont {Guerin}}, \bibinfo {author} {\bibfnamefont
  {S.}~\bibnamefont {Habegger}}, \bibinfo {author} {\bibfnamefont {M.~P.}\
  \bibnamefont {Harrigan}}, \bibinfo {author} {\bibfnamefont {M.~J.}\
  \bibnamefont {Hartmann}}, \bibinfo {author} {\bibfnamefont {A.}~\bibnamefont
  {Ho}}, \bibinfo {author} {\bibfnamefont {M.}~\bibnamefont {Hoffmann}},
  \bibinfo {author} {\bibfnamefont {T.}~\bibnamefont {Huang}}, \bibinfo
  {author} {\bibfnamefont {T.~S.}\ \bibnamefont {Humble}}, \bibinfo {author}
  {\bibfnamefont {S.~V.}\ \bibnamefont {Isakov}}, \bibinfo {author}
  {\bibfnamefont {E.}~\bibnamefont {Jeffrey}}, \bibinfo {author} {\bibfnamefont
  {Z.}~\bibnamefont {Jiang}}, \bibinfo {author} {\bibfnamefont
  {D.}~\bibnamefont {Kafri}}, \bibinfo {author} {\bibfnamefont
  {K.}~\bibnamefont {Kechedzhi}}, \bibinfo {author} {\bibfnamefont
  {J.}~\bibnamefont {Kelly}}, \bibinfo {author} {\bibfnamefont {P.~V.}\
  \bibnamefont {Klimov}}, \bibinfo {author} {\bibfnamefont {S.}~\bibnamefont
  {Knysh}}, \bibinfo {author} {\bibfnamefont {A.}~\bibnamefont {Korotkov}},
  \bibinfo {author} {\bibfnamefont {F.}~\bibnamefont {Kostritsa}}, \bibinfo
  {author} {\bibfnamefont {D.}~\bibnamefont {Landhuis}}, \bibinfo {author}
  {\bibfnamefont {M.}~\bibnamefont {Lindmark}}, \bibinfo {author}
  {\bibfnamefont {E.}~\bibnamefont {Lucero}}, \bibinfo {author} {\bibfnamefont
  {D.}~\bibnamefont {Lyakh}}, \bibinfo {author} {\bibfnamefont
  {S.}~\bibnamefont {Mandrà}}, \bibinfo {author} {\bibfnamefont {J.~R.}\
  \bibnamefont {McClean}}, \bibinfo {author} {\bibfnamefont {M.}~\bibnamefont
  {McEwen}}, \bibinfo {author} {\bibfnamefont {A.}~\bibnamefont {Megrant}},
  \bibinfo {author} {\bibfnamefont {X.}~\bibnamefont {Mi}}, \bibinfo {author}
  {\bibfnamefont {K.}~\bibnamefont {Michielsen}}, \bibinfo {author}
  {\bibfnamefont {M.}~\bibnamefont {Mohseni}}, \bibinfo {author} {\bibfnamefont
  {J.}~\bibnamefont {Mutus}}, \bibinfo {author} {\bibfnamefont
  {O.}~\bibnamefont {Naaman}}, \bibinfo {author} {\bibfnamefont
  {M.}~\bibnamefont {Neeley}}, \bibinfo {author} {\bibfnamefont
  {C.}~\bibnamefont {Neill}}, \bibinfo {author} {\bibfnamefont {M.~Y.}\
  \bibnamefont {Niu}}, \bibinfo {author} {\bibfnamefont {E.}~\bibnamefont
  {Ostby}}, \bibinfo {author} {\bibfnamefont {A.}~\bibnamefont {Petukhov}},
  \bibinfo {author} {\bibfnamefont {J.~C.}\ \bibnamefont {Platt}}, \bibinfo
  {author} {\bibfnamefont {C.}~\bibnamefont {Quintana}}, \bibinfo {author}
  {\bibfnamefont {E.~G.}\ \bibnamefont {Rieffel}}, \bibinfo {author}
  {\bibfnamefont {P.}~\bibnamefont {Roushan}}, \bibinfo {author} {\bibfnamefont
  {N.~C.}\ \bibnamefont {Rubin}}, \bibinfo {author} {\bibfnamefont
  {D.}~\bibnamefont {Sank}}, \bibinfo {author} {\bibfnamefont {K.~J.}\
  \bibnamefont {Satzinger}}, \bibinfo {author} {\bibfnamefont {V.}~\bibnamefont
  {Smelyanskiy}}, \bibinfo {author} {\bibfnamefont {K.~J.}\ \bibnamefont
  {Sung}}, \bibinfo {author} {\bibfnamefont {M.~D.}\ \bibnamefont
  {Trevithick}}, \bibinfo {author} {\bibfnamefont {A.}~\bibnamefont
  {Vainsencher}}, \bibinfo {author} {\bibfnamefont {B.}~\bibnamefont
  {Villalonga}}, \bibinfo {author} {\bibfnamefont {T.}~\bibnamefont {White}},
  \bibinfo {author} {\bibfnamefont {Z.~J.}\ \bibnamefont {Yao}}, \bibinfo
  {author} {\bibfnamefont {P.}~\bibnamefont {Yeh}}, \bibinfo {author}
  {\bibfnamefont {A.}~\bibnamefont {Zalcman}}, \bibinfo {author} {\bibfnamefont
  {H.}~\bibnamefont {Neven}}, \ and\ \bibinfo {author} {\bibfnamefont {J.~M.}\
  \bibnamefont {Martinis}},\ }\href {\doibase 10.1038/s41586-019-1666-5}
  {\bibfield  {journal} {\bibinfo  {journal} {Nature}\ }\textbf {\bibinfo
  {volume} {574}},\ \bibinfo {pages} {505} (\bibinfo {year}
  {2019})}\BibitemShut {NoStop}%
\bibitem [{\citenamefont {Wu}\ \emph {et~al.}(2021)\citenamefont {Wu},
  \citenamefont {Bao}, \citenamefont {Cao}, \citenamefont {Chen}, \citenamefont
  {Chen}, \citenamefont {Chen}, \citenamefont {Chung}, \citenamefont {Deng},
  \citenamefont {Du}, \citenamefont {Fan}, \citenamefont {Gong}, \citenamefont
  {Guo}, \citenamefont {Guo}, \citenamefont {Guo}, \citenamefont {Han},
  \citenamefont {Hong}, \citenamefont {Huang}, \citenamefont {Huo},
  \citenamefont {Li}, \citenamefont {Li}, \citenamefont {Li}, \citenamefont
  {Li}, \citenamefont {Liang}, \citenamefont {Lin}, \citenamefont {Lin},
  \citenamefont {Qian}, \citenamefont {Qiao}, \citenamefont {Rong},
  \citenamefont {Su}, \citenamefont {Sun}, \citenamefont {Wang}, \citenamefont
  {Wang}, \citenamefont {Wu}, \citenamefont {Xu}, \citenamefont {Yan},
  \citenamefont {Yang}, \citenamefont {Yang}, \citenamefont {Ye}, \citenamefont
  {Yin}, \citenamefont {Ying}, \citenamefont {Yu}, \citenamefont {Zha},
  \citenamefont {Zhang}, \citenamefont {Zhang}, \citenamefont {Zhang},
  \citenamefont {Zhang}, \citenamefont {Zhao}, \citenamefont {Zhao},
  \citenamefont {Zhou}, \citenamefont {Zhu}, \citenamefont {Lu}, \citenamefont
  {Peng}, \citenamefont {Zhu},\ and\ \citenamefont {Pan}}]{wu_strong_2021}%
  \BibitemOpen
  \bibfield  {author} {\bibinfo {author} {\bibfnamefont {Y.}~\bibnamefont
  {Wu}}, \bibinfo {author} {\bibfnamefont {W.-S.}\ \bibnamefont {Bao}},
  \bibinfo {author} {\bibfnamefont {S.}~\bibnamefont {Cao}}, \bibinfo {author}
  {\bibfnamefont {F.}~\bibnamefont {Chen}}, \bibinfo {author} {\bibfnamefont
  {M.-C.}\ \bibnamefont {Chen}}, \bibinfo {author} {\bibfnamefont
  {X.}~\bibnamefont {Chen}}, \bibinfo {author} {\bibfnamefont {T.-H.}\
  \bibnamefont {Chung}}, \bibinfo {author} {\bibfnamefont {H.}~\bibnamefont
  {Deng}}, \bibinfo {author} {\bibfnamefont {Y.}~\bibnamefont {Du}}, \bibinfo
  {author} {\bibfnamefont {D.}~\bibnamefont {Fan}}, \bibinfo {author}
  {\bibfnamefont {M.}~\bibnamefont {Gong}}, \bibinfo {author} {\bibfnamefont
  {C.}~\bibnamefont {Guo}}, \bibinfo {author} {\bibfnamefont {C.}~\bibnamefont
  {Guo}}, \bibinfo {author} {\bibfnamefont {S.}~\bibnamefont {Guo}}, \bibinfo
  {author} {\bibfnamefont {L.}~\bibnamefont {Han}}, \bibinfo {author}
  {\bibfnamefont {L.}~\bibnamefont {Hong}}, \bibinfo {author} {\bibfnamefont
  {H.-L.}\ \bibnamefont {Huang}}, \bibinfo {author} {\bibfnamefont {Y.-H.}\
  \bibnamefont {Huo}}, \bibinfo {author} {\bibfnamefont {L.}~\bibnamefont
  {Li}}, \bibinfo {author} {\bibfnamefont {N.}~\bibnamefont {Li}}, \bibinfo
  {author} {\bibfnamefont {S.}~\bibnamefont {Li}}, \bibinfo {author}
  {\bibfnamefont {Y.}~\bibnamefont {Li}}, \bibinfo {author} {\bibfnamefont
  {F.}~\bibnamefont {Liang}}, \bibinfo {author} {\bibfnamefont
  {C.}~\bibnamefont {Lin}}, \bibinfo {author} {\bibfnamefont {J.}~\bibnamefont
  {Lin}}, \bibinfo {author} {\bibfnamefont {H.}~\bibnamefont {Qian}}, \bibinfo
  {author} {\bibfnamefont {D.}~\bibnamefont {Qiao}}, \bibinfo {author}
  {\bibfnamefont {H.}~\bibnamefont {Rong}}, \bibinfo {author} {\bibfnamefont
  {H.}~\bibnamefont {Su}}, \bibinfo {author} {\bibfnamefont {L.}~\bibnamefont
  {Sun}}, \bibinfo {author} {\bibfnamefont {L.}~\bibnamefont {Wang}}, \bibinfo
  {author} {\bibfnamefont {S.}~\bibnamefont {Wang}}, \bibinfo {author}
  {\bibfnamefont {D.}~\bibnamefont {Wu}}, \bibinfo {author} {\bibfnamefont
  {Y.}~\bibnamefont {Xu}}, \bibinfo {author} {\bibfnamefont {K.}~\bibnamefont
  {Yan}}, \bibinfo {author} {\bibfnamefont {W.}~\bibnamefont {Yang}}, \bibinfo
  {author} {\bibfnamefont {Y.}~\bibnamefont {Yang}}, \bibinfo {author}
  {\bibfnamefont {Y.}~\bibnamefont {Ye}}, \bibinfo {author} {\bibfnamefont
  {J.}~\bibnamefont {Yin}}, \bibinfo {author} {\bibfnamefont {C.}~\bibnamefont
  {Ying}}, \bibinfo {author} {\bibfnamefont {J.}~\bibnamefont {Yu}}, \bibinfo
  {author} {\bibfnamefont {C.}~\bibnamefont {Zha}}, \bibinfo {author}
  {\bibfnamefont {C.}~\bibnamefont {Zhang}}, \bibinfo {author} {\bibfnamefont
  {H.}~\bibnamefont {Zhang}}, \bibinfo {author} {\bibfnamefont
  {K.}~\bibnamefont {Zhang}}, \bibinfo {author} {\bibfnamefont
  {Y.}~\bibnamefont {Zhang}}, \bibinfo {author} {\bibfnamefont
  {H.}~\bibnamefont {Zhao}}, \bibinfo {author} {\bibfnamefont {Y.}~\bibnamefont
  {Zhao}}, \bibinfo {author} {\bibfnamefont {L.}~\bibnamefont {Zhou}}, \bibinfo
  {author} {\bibfnamefont {Q.}~\bibnamefont {Zhu}}, \bibinfo {author}
  {\bibfnamefont {C.-Y.}\ \bibnamefont {Lu}}, \bibinfo {author} {\bibfnamefont
  {C.-Z.}\ \bibnamefont {Peng}}, \bibinfo {author} {\bibfnamefont
  {X.}~\bibnamefont {Zhu}}, \ and\ \bibinfo {author} {\bibfnamefont {J.-W.}\
  \bibnamefont {Pan}},\ }\href {\doibase 10.1103/PhysRevLett.127.180501}
  {\bibfield  {journal} {\bibinfo  {journal} {Phys. Rev. Lett.}\ }\textbf
  {\bibinfo {volume} {127}},\ \bibinfo {pages} {180501} (\bibinfo {year}
  {2021})}\BibitemShut {NoStop}%
\bibitem [{\citenamefont {Huang}\ \emph {et~al.}(2020)\citenamefont {Huang},
  \citenamefont {Wu}, \citenamefont {Fan},\ and\ \citenamefont
  {Zhu}}]{huang_superconducting_2020}%
  \BibitemOpen
  \bibfield  {author} {\bibinfo {author} {\bibfnamefont {H.-L.}\ \bibnamefont
  {Huang}}, \bibinfo {author} {\bibfnamefont {D.}~\bibnamefont {Wu}}, \bibinfo
  {author} {\bibfnamefont {D.}~\bibnamefont {Fan}}, \ and\ \bibinfo {author}
  {\bibfnamefont {X.}~\bibnamefont {Zhu}},\ }\href {\doibase
  10.1007/s11432-020-2881-9} {\bibfield  {journal} {\bibinfo  {journal} {Sci.
  China Inf. Sci.}\ }\textbf {\bibinfo {volume} {63}},\ \bibinfo {pages}
  {180501} (\bibinfo {year} {2020})}\BibitemShut {NoStop}%
\bibitem [{\citenamefont {Gong}\ \emph {et~al.}(2021)\citenamefont {Gong},
  \citenamefont {Wang}, \citenamefont {Zha}, \citenamefont {Chen},
  \citenamefont {Huang}, \citenamefont {Wu}, \citenamefont {Zhu}, \citenamefont
  {Zhao}, \citenamefont {Li}, \citenamefont {Guo}, \citenamefont {Qian},
  \citenamefont {Ye}, \citenamefont {Chen}, \citenamefont {Ying}, \citenamefont
  {Yu}, \citenamefont {Fan}, \citenamefont {Wu}, \citenamefont {Su},
  \citenamefont {Deng}, \citenamefont {Rong}, \citenamefont {Zhang},
  \citenamefont {Cao}, \citenamefont {Lin}, \citenamefont {Xu}, \citenamefont
  {Sun}, \citenamefont {Guo}, \citenamefont {Li}, \citenamefont {Liang},
  \citenamefont {Bastidas}, \citenamefont {Nemoto}, \citenamefont {Munro},
  \citenamefont {Huo}, \citenamefont {Lu}, \citenamefont {Peng}, \citenamefont
  {Zhu},\ and\ \citenamefont {Pan}}]{gong_quantum_2021}%
  \BibitemOpen
  \bibfield  {author} {\bibinfo {author} {\bibfnamefont {M.}~\bibnamefont
  {Gong}}, \bibinfo {author} {\bibfnamefont {S.}~\bibnamefont {Wang}}, \bibinfo
  {author} {\bibfnamefont {C.}~\bibnamefont {Zha}}, \bibinfo {author}
  {\bibfnamefont {M.-C.}\ \bibnamefont {Chen}}, \bibinfo {author}
  {\bibfnamefont {H.-L.}\ \bibnamefont {Huang}}, \bibinfo {author}
  {\bibfnamefont {Y.}~\bibnamefont {Wu}}, \bibinfo {author} {\bibfnamefont
  {Q.}~\bibnamefont {Zhu}}, \bibinfo {author} {\bibfnamefont {Y.}~\bibnamefont
  {Zhao}}, \bibinfo {author} {\bibfnamefont {S.}~\bibnamefont {Li}}, \bibinfo
  {author} {\bibfnamefont {S.}~\bibnamefont {Guo}}, \bibinfo {author}
  {\bibfnamefont {H.}~\bibnamefont {Qian}}, \bibinfo {author} {\bibfnamefont
  {Y.}~\bibnamefont {Ye}}, \bibinfo {author} {\bibfnamefont {F.}~\bibnamefont
  {Chen}}, \bibinfo {author} {\bibfnamefont {C.}~\bibnamefont {Ying}}, \bibinfo
  {author} {\bibfnamefont {J.}~\bibnamefont {Yu}}, \bibinfo {author}
  {\bibfnamefont {D.}~\bibnamefont {Fan}}, \bibinfo {author} {\bibfnamefont
  {D.}~\bibnamefont {Wu}}, \bibinfo {author} {\bibfnamefont {H.}~\bibnamefont
  {Su}}, \bibinfo {author} {\bibfnamefont {H.}~\bibnamefont {Deng}}, \bibinfo
  {author} {\bibfnamefont {H.}~\bibnamefont {Rong}}, \bibinfo {author}
  {\bibfnamefont {K.}~\bibnamefont {Zhang}}, \bibinfo {author} {\bibfnamefont
  {S.}~\bibnamefont {Cao}}, \bibinfo {author} {\bibfnamefont {J.}~\bibnamefont
  {Lin}}, \bibinfo {author} {\bibfnamefont {Y.}~\bibnamefont {Xu}}, \bibinfo
  {author} {\bibfnamefont {L.}~\bibnamefont {Sun}}, \bibinfo {author}
  {\bibfnamefont {C.}~\bibnamefont {Guo}}, \bibinfo {author} {\bibfnamefont
  {N.}~\bibnamefont {Li}}, \bibinfo {author} {\bibfnamefont {F.}~\bibnamefont
  {Liang}}, \bibinfo {author} {\bibfnamefont {V.~M.}\ \bibnamefont {Bastidas}},
  \bibinfo {author} {\bibfnamefont {K.}~\bibnamefont {Nemoto}}, \bibinfo
  {author} {\bibfnamefont {W.~J.}\ \bibnamefont {Munro}}, \bibinfo {author}
  {\bibfnamefont {Y.-H.}\ \bibnamefont {Huo}}, \bibinfo {author} {\bibfnamefont
  {C.-Y.}\ \bibnamefont {Lu}}, \bibinfo {author} {\bibfnamefont {C.-Z.}\
  \bibnamefont {Peng}}, \bibinfo {author} {\bibfnamefont {X.}~\bibnamefont
  {Zhu}}, \ and\ \bibinfo {author} {\bibfnamefont {J.-W.}\ \bibnamefont
  {Pan}},\ }\href {\doibase 10.1126/science.abg7812} {\bibfield  {journal}
  {\bibinfo  {journal} {Science}\ }\textbf {\bibinfo {volume} {372}},\ \bibinfo
  {pages} {948} (\bibinfo {year} {2021})}\BibitemShut {NoStop}%
\bibitem [{\citenamefont {Bruzewicz}\ \emph {et~al.}(2019)\citenamefont
  {Bruzewicz}, \citenamefont {Chiaverini}, \citenamefont {McConnell},\ and\
  \citenamefont {Sage}}]{bruzewicz_trapped-ion_2019}%
  \BibitemOpen
  \bibfield  {author} {\bibinfo {author} {\bibfnamefont {C.~D.}\ \bibnamefont
  {Bruzewicz}}, \bibinfo {author} {\bibfnamefont {J.}~\bibnamefont
  {Chiaverini}}, \bibinfo {author} {\bibfnamefont {R.}~\bibnamefont
  {McConnell}}, \ and\ \bibinfo {author} {\bibfnamefont {J.~M.}\ \bibnamefont
  {Sage}},\ }\href {\doibase 10.1063/1.5088164} {\bibfield  {journal} {\bibinfo
   {journal} {Applied Physics Reviews}\ }\textbf {\bibinfo {volume} {6}},\
  \bibinfo {pages} {021314} (\bibinfo {year} {2019})}\BibitemShut {NoStop}%
\bibitem [{\citenamefont {Zhong}\ \emph {et~al.}(2020)\citenamefont {Zhong},
  \citenamefont {Wang}, \citenamefont {Deng}, \citenamefont {Chen},
  \citenamefont {Peng}, \citenamefont {Luo}, \citenamefont {Qin}, \citenamefont
  {Wu}, \citenamefont {Ding}, \citenamefont {Hu}, \citenamefont {Hu},
  \citenamefont {Yang}, \citenamefont {Zhang}, \citenamefont {Li},
  \citenamefont {Li}, \citenamefont {Jiang}, \citenamefont {Gan}, \citenamefont
  {Yang}, \citenamefont {You}, \citenamefont {Wang}, \citenamefont {Li},
  \citenamefont {Liu}, \citenamefont {Lu},\ and\ \citenamefont
  {Pan}}]{zhong_quantum_2020}%
  \BibitemOpen
  \bibfield  {author} {\bibinfo {author} {\bibfnamefont {H.-S.}\ \bibnamefont
  {Zhong}}, \bibinfo {author} {\bibfnamefont {H.}~\bibnamefont {Wang}},
  \bibinfo {author} {\bibfnamefont {Y.-H.}\ \bibnamefont {Deng}}, \bibinfo
  {author} {\bibfnamefont {M.-C.}\ \bibnamefont {Chen}}, \bibinfo {author}
  {\bibfnamefont {L.-C.}\ \bibnamefont {Peng}}, \bibinfo {author}
  {\bibfnamefont {Y.-H.}\ \bibnamefont {Luo}}, \bibinfo {author} {\bibfnamefont
  {J.}~\bibnamefont {Qin}}, \bibinfo {author} {\bibfnamefont {D.}~\bibnamefont
  {Wu}}, \bibinfo {author} {\bibfnamefont {X.}~\bibnamefont {Ding}}, \bibinfo
  {author} {\bibfnamefont {Y.}~\bibnamefont {Hu}}, \bibinfo {author}
  {\bibfnamefont {P.}~\bibnamefont {Hu}}, \bibinfo {author} {\bibfnamefont
  {X.-Y.}\ \bibnamefont {Yang}}, \bibinfo {author} {\bibfnamefont {W.-J.}\
  \bibnamefont {Zhang}}, \bibinfo {author} {\bibfnamefont {H.}~\bibnamefont
  {Li}}, \bibinfo {author} {\bibfnamefont {Y.}~\bibnamefont {Li}}, \bibinfo
  {author} {\bibfnamefont {X.}~\bibnamefont {Jiang}}, \bibinfo {author}
  {\bibfnamefont {L.}~\bibnamefont {Gan}}, \bibinfo {author} {\bibfnamefont
  {G.}~\bibnamefont {Yang}}, \bibinfo {author} {\bibfnamefont {L.}~\bibnamefont
  {You}}, \bibinfo {author} {\bibfnamefont {Z.}~\bibnamefont {Wang}}, \bibinfo
  {author} {\bibfnamefont {L.}~\bibnamefont {Li}}, \bibinfo {author}
  {\bibfnamefont {N.-L.}\ \bibnamefont {Liu}}, \bibinfo {author} {\bibfnamefont
  {C.-Y.}\ \bibnamefont {Lu}}, \ and\ \bibinfo {author} {\bibfnamefont {J.-W.}\
  \bibnamefont {Pan}},\ }\href {\doibase 10.1126/science.abe8770} {\bibfield
  {journal} {\bibinfo  {journal} {Science}\ }\textbf {\bibinfo {volume}
  {370}},\ \bibinfo {pages} {1460} (\bibinfo {year} {2020})}\BibitemShut
  {NoStop}%
\bibitem [{\citenamefont {Arrazola}\ \emph {et~al.}(2021)\citenamefont
  {Arrazola}, \citenamefont {Bergholm}, \citenamefont {Brádler}, \citenamefont
  {Bromley}, \citenamefont {Collins}, \citenamefont {Dhand}, \citenamefont
  {Fumagalli}, \citenamefont {Gerrits}, \citenamefont {Goussev}, \citenamefont
  {Helt}, \citenamefont {Hundal}, \citenamefont {Isacsson}, \citenamefont
  {Israel}, \citenamefont {Izaac}, \citenamefont {Jahangiri}, \citenamefont
  {Janik}, \citenamefont {Killoran}, \citenamefont {Kumar}, \citenamefont
  {Lavoie}, \citenamefont {Lita}, \citenamefont {Mahler}, \citenamefont
  {Menotti}, \citenamefont {Morrison}, \citenamefont {Nam}, \citenamefont
  {Neuhaus}, \citenamefont {Qi}, \citenamefont {Quesada}, \citenamefont
  {Repingon}, \citenamefont {Sabapathy}, \citenamefont {Schuld}, \citenamefont
  {Su}, \citenamefont {Swinarton}, \citenamefont {Száva}, \citenamefont {Tan},
  \citenamefont {Tan}, \citenamefont {Vaidya}, \citenamefont {Vernon},
  \citenamefont {Zabaneh},\ and\ \citenamefont
  {Zhang}}]{arrazola_quantum_2021}%
  \BibitemOpen
  \bibfield  {author} {\bibinfo {author} {\bibfnamefont {J.~M.}\ \bibnamefont
  {Arrazola}}, \bibinfo {author} {\bibfnamefont {V.}~\bibnamefont {Bergholm}},
  \bibinfo {author} {\bibfnamefont {K.}~\bibnamefont {Brádler}}, \bibinfo
  {author} {\bibfnamefont {T.~R.}\ \bibnamefont {Bromley}}, \bibinfo {author}
  {\bibfnamefont {M.~J.}\ \bibnamefont {Collins}}, \bibinfo {author}
  {\bibfnamefont {I.}~\bibnamefont {Dhand}}, \bibinfo {author} {\bibfnamefont
  {A.}~\bibnamefont {Fumagalli}}, \bibinfo {author} {\bibfnamefont
  {T.}~\bibnamefont {Gerrits}}, \bibinfo {author} {\bibfnamefont
  {A.}~\bibnamefont {Goussev}}, \bibinfo {author} {\bibfnamefont {L.~G.}\
  \bibnamefont {Helt}}, \bibinfo {author} {\bibfnamefont {J.}~\bibnamefont
  {Hundal}}, \bibinfo {author} {\bibfnamefont {T.}~\bibnamefont {Isacsson}},
  \bibinfo {author} {\bibfnamefont {R.~B.}\ \bibnamefont {Israel}}, \bibinfo
  {author} {\bibfnamefont {J.}~\bibnamefont {Izaac}}, \bibinfo {author}
  {\bibfnamefont {S.}~\bibnamefont {Jahangiri}}, \bibinfo {author}
  {\bibfnamefont {R.}~\bibnamefont {Janik}}, \bibinfo {author} {\bibfnamefont
  {N.}~\bibnamefont {Killoran}}, \bibinfo {author} {\bibfnamefont {S.~P.}\
  \bibnamefont {Kumar}}, \bibinfo {author} {\bibfnamefont {J.}~\bibnamefont
  {Lavoie}}, \bibinfo {author} {\bibfnamefont {A.~E.}\ \bibnamefont {Lita}},
  \bibinfo {author} {\bibfnamefont {D.~H.}\ \bibnamefont {Mahler}}, \bibinfo
  {author} {\bibfnamefont {M.}~\bibnamefont {Menotti}}, \bibinfo {author}
  {\bibfnamefont {B.}~\bibnamefont {Morrison}}, \bibinfo {author}
  {\bibfnamefont {S.~W.}\ \bibnamefont {Nam}}, \bibinfo {author} {\bibfnamefont
  {L.}~\bibnamefont {Neuhaus}}, \bibinfo {author} {\bibfnamefont {H.~Y.}\
  \bibnamefont {Qi}}, \bibinfo {author} {\bibfnamefont {N.}~\bibnamefont
  {Quesada}}, \bibinfo {author} {\bibfnamefont {A.}~\bibnamefont {Repingon}},
  \bibinfo {author} {\bibfnamefont {K.~K.}\ \bibnamefont {Sabapathy}}, \bibinfo
  {author} {\bibfnamefont {M.}~\bibnamefont {Schuld}}, \bibinfo {author}
  {\bibfnamefont {D.}~\bibnamefont {Su}}, \bibinfo {author} {\bibfnamefont
  {J.}~\bibnamefont {Swinarton}}, \bibinfo {author} {\bibfnamefont
  {A.}~\bibnamefont {Száva}}, \bibinfo {author} {\bibfnamefont
  {K.}~\bibnamefont {Tan}}, \bibinfo {author} {\bibfnamefont {P.}~\bibnamefont
  {Tan}}, \bibinfo {author} {\bibfnamefont {V.~D.}\ \bibnamefont {Vaidya}},
  \bibinfo {author} {\bibfnamefont {Z.}~\bibnamefont {Vernon}}, \bibinfo
  {author} {\bibfnamefont {Z.}~\bibnamefont {Zabaneh}}, \ and\ \bibinfo
  {author} {\bibfnamefont {Y.}~\bibnamefont {Zhang}},\ }\href {\doibase
  10.1038/s41586-021-03202-1} {\bibfield  {journal} {\bibinfo  {journal}
  {Nature}\ }\textbf {\bibinfo {volume} {591}},\ \bibinfo {pages} {54}
  (\bibinfo {year} {2021})}\BibitemShut {NoStop}%
\bibitem [{\citenamefont {Zhong}\ \emph {et~al.}(2021)\citenamefont {Zhong},
  \citenamefont {Deng}, \citenamefont {Qin}, \citenamefont {Wang},
  \citenamefont {Chen}, \citenamefont {Peng}, \citenamefont {Luo},
  \citenamefont {Wu}, \citenamefont {Gong}, \citenamefont {Su}, \citenamefont
  {Hu}, \citenamefont {Hu}, \citenamefont {Yang}, \citenamefont {Zhang},
  \citenamefont {Li}, \citenamefont {Li}, \citenamefont {Jiang}, \citenamefont
  {Gan}, \citenamefont {Yang}, \citenamefont {You}, \citenamefont {Wang},
  \citenamefont {Li}, \citenamefont {Liu}, \citenamefont {Renema},
  \citenamefont {Lu},\ and\ \citenamefont
  {Pan}}]{zhong_phase-programmable_2021}%
  \BibitemOpen
  \bibfield  {author} {\bibinfo {author} {\bibfnamefont {H.-S.}\ \bibnamefont
  {Zhong}}, \bibinfo {author} {\bibfnamefont {Y.-H.}\ \bibnamefont {Deng}},
  \bibinfo {author} {\bibfnamefont {J.}~\bibnamefont {Qin}}, \bibinfo {author}
  {\bibfnamefont {H.}~\bibnamefont {Wang}}, \bibinfo {author} {\bibfnamefont
  {M.-C.}\ \bibnamefont {Chen}}, \bibinfo {author} {\bibfnamefont {L.-C.}\
  \bibnamefont {Peng}}, \bibinfo {author} {\bibfnamefont {Y.-H.}\ \bibnamefont
  {Luo}}, \bibinfo {author} {\bibfnamefont {D.}~\bibnamefont {Wu}}, \bibinfo
  {author} {\bibfnamefont {S.-Q.}\ \bibnamefont {Gong}}, \bibinfo {author}
  {\bibfnamefont {H.}~\bibnamefont {Su}}, \bibinfo {author} {\bibfnamefont
  {Y.}~\bibnamefont {Hu}}, \bibinfo {author} {\bibfnamefont {P.}~\bibnamefont
  {Hu}}, \bibinfo {author} {\bibfnamefont {X.-Y.}\ \bibnamefont {Yang}},
  \bibinfo {author} {\bibfnamefont {W.-J.}\ \bibnamefont {Zhang}}, \bibinfo
  {author} {\bibfnamefont {H.}~\bibnamefont {Li}}, \bibinfo {author}
  {\bibfnamefont {Y.}~\bibnamefont {Li}}, \bibinfo {author} {\bibfnamefont
  {X.}~\bibnamefont {Jiang}}, \bibinfo {author} {\bibfnamefont
  {L.}~\bibnamefont {Gan}}, \bibinfo {author} {\bibfnamefont {G.}~\bibnamefont
  {Yang}}, \bibinfo {author} {\bibfnamefont {L.}~\bibnamefont {You}}, \bibinfo
  {author} {\bibfnamefont {Z.}~\bibnamefont {Wang}}, \bibinfo {author}
  {\bibfnamefont {L.}~\bibnamefont {Li}}, \bibinfo {author} {\bibfnamefont
  {N.-L.}\ \bibnamefont {Liu}}, \bibinfo {author} {\bibfnamefont {J.~J.}\
  \bibnamefont {Renema}}, \bibinfo {author} {\bibfnamefont {C.-Y.}\
  \bibnamefont {Lu}}, \ and\ \bibinfo {author} {\bibfnamefont {J.-W.}\
  \bibnamefont {Pan}},\ }\href {\doibase 10.1103/PhysRevLett.127.180502}
  {\bibfield  {journal} {\bibinfo  {journal} {Phys. Rev. Lett.}\ }\textbf
  {\bibinfo {volume} {127}},\ \bibinfo {pages} {180502} (\bibinfo {year}
  {2021})}\BibitemShut {NoStop}%
\bibitem [{\citenamefont {Slussarenko}\ and\ \citenamefont
  {Pryde}(2019)}]{slussarenko_photonic_2019}%
  \BibitemOpen
  \bibfield  {author} {\bibinfo {author} {\bibfnamefont {S.}~\bibnamefont
  {Slussarenko}}\ and\ \bibinfo {author} {\bibfnamefont {G.~J.}\ \bibnamefont
  {Pryde}},\ }\href {\doibase 10.1063/1.5115814} {\bibfield  {journal}
  {\bibinfo  {journal} {Applied Physics Reviews}\ }\textbf {\bibinfo {volume}
  {6}},\ \bibinfo {pages} {041303} (\bibinfo {year} {2019})}\BibitemShut
  {NoStop}%
\bibitem [{\citenamefont {Bourassa}\ \emph {et~al.}(2021)\citenamefont
  {Bourassa}, \citenamefont {Alexander}, \citenamefont {Vasmer}, \citenamefont
  {Patil}, \citenamefont {Tzitrin}, \citenamefont {Matsuura}, \citenamefont
  {Su}, \citenamefont {Baragiola}, \citenamefont {Guha}, \citenamefont
  {Dauphinais}, \citenamefont {Sabapathy}, \citenamefont {Menicucci},\ and\
  \citenamefont {Dhand}}]{bourassa_blueprint_2021}%
  \BibitemOpen
  \bibfield  {author} {\bibinfo {author} {\bibfnamefont {J.~E.}\ \bibnamefont
  {Bourassa}}, \bibinfo {author} {\bibfnamefont {R.~N.}\ \bibnamefont
  {Alexander}}, \bibinfo {author} {\bibfnamefont {M.}~\bibnamefont {Vasmer}},
  \bibinfo {author} {\bibfnamefont {A.}~\bibnamefont {Patil}}, \bibinfo
  {author} {\bibfnamefont {I.}~\bibnamefont {Tzitrin}}, \bibinfo {author}
  {\bibfnamefont {T.}~\bibnamefont {Matsuura}}, \bibinfo {author}
  {\bibfnamefont {D.}~\bibnamefont {Su}}, \bibinfo {author} {\bibfnamefont
  {B.~Q.}\ \bibnamefont {Baragiola}}, \bibinfo {author} {\bibfnamefont
  {S.}~\bibnamefont {Guha}}, \bibinfo {author} {\bibfnamefont {G.}~\bibnamefont
  {Dauphinais}}, \bibinfo {author} {\bibfnamefont {K.~K.}\ \bibnamefont
  {Sabapathy}}, \bibinfo {author} {\bibfnamefont {N.~C.}\ \bibnamefont
  {Menicucci}}, \ and\ \bibinfo {author} {\bibfnamefont {I.}~\bibnamefont
  {Dhand}},\ }\href {\doibase 10.22331/q-2021-02-04-392} {\bibfield  {journal}
  {\bibinfo  {journal} {Quantum}\ }\textbf {\bibinfo {volume} {5}},\ \bibinfo
  {pages} {392} (\bibinfo {year} {2021})}\BibitemShut {NoStop}%
\bibitem [{\citenamefont {McClean}\ \emph {et~al.}(2016)\citenamefont
  {McClean}, \citenamefont {Romero}, \citenamefont {Babbush},\ and\
  \citenamefont {Aspuru-Guzik}}]{McClean_2016}%
  \BibitemOpen
  \bibfield  {author} {\bibinfo {author} {\bibfnamefont {J.~R.}\ \bibnamefont
  {McClean}}, \bibinfo {author} {\bibfnamefont {J.}~\bibnamefont {Romero}},
  \bibinfo {author} {\bibfnamefont {R.}~\bibnamefont {Babbush}}, \ and\
  \bibinfo {author} {\bibfnamefont {A.}~\bibnamefont {Aspuru-Guzik}},\ }\href
  {\doibase 10.1088/1367-2630/18/2/023023} {\bibfield  {journal} {\bibinfo
  {journal} {New Journal of Physics}\ }\textbf {\bibinfo {volume} {18}},\
  \bibinfo {pages} {023023} (\bibinfo {year} {2016})}\BibitemShut {NoStop}%
\bibitem [{\citenamefont {Farhi}\ \emph {et~al.}(2014)\citenamefont {Farhi},
  \citenamefont {Goldstone},\ and\ \citenamefont
  {Gutmann}}]{farhi_quantum_2014}%
  \BibitemOpen
  \bibfield  {author} {\bibinfo {author} {\bibfnamefont {E.}~\bibnamefont
  {Farhi}}, \bibinfo {author} {\bibfnamefont {J.}~\bibnamefont {Goldstone}}, \
  and\ \bibinfo {author} {\bibfnamefont {S.}~\bibnamefont {Gutmann}},\ }\href
  {http://arxiv.org/abs/1411.4028} {\bibfield  {journal} {\bibinfo  {journal}
  {arXiv:1411.4028 [quant-ph]}\ } (\bibinfo {year} {2014})}\BibitemShut
  {NoStop}%
\bibitem [{\citenamefont {Bouland}\ \emph {et~al.}(2019)\citenamefont
  {Bouland}, \citenamefont {Fefferman}, \citenamefont {Nirkhe},\ and\
  \citenamefont {Vazirani}}]{bouland_complexity_2019}%
  \BibitemOpen
  \bibfield  {author} {\bibinfo {author} {\bibfnamefont {A.}~\bibnamefont
  {Bouland}}, \bibinfo {author} {\bibfnamefont {B.}~\bibnamefont {Fefferman}},
  \bibinfo {author} {\bibfnamefont {C.}~\bibnamefont {Nirkhe}}, \ and\ \bibinfo
  {author} {\bibfnamefont {U.}~\bibnamefont {Vazirani}},\ }\href {\doibase
  10.1038/s41567-018-0318-2} {\bibfield  {journal} {\bibinfo  {journal} {Nature
  Phys}\ }\textbf {\bibinfo {volume} {15}},\ \bibinfo {pages} {159} (\bibinfo
  {year} {2019})}\BibitemShut {NoStop}%
\bibitem [{\citenamefont {Rahimi-Keshari}\ \emph {et~al.}(2015)\citenamefont
  {Rahimi-Keshari}, \citenamefont {Lund},\ and\ \citenamefont
  {Ralph}}]{rahimi-keshari_what_2015}%
  \BibitemOpen
  \bibfield  {author} {\bibinfo {author} {\bibfnamefont {S.}~\bibnamefont
  {Rahimi-Keshari}}, \bibinfo {author} {\bibfnamefont {A.~P.}\ \bibnamefont
  {Lund}}, \ and\ \bibinfo {author} {\bibfnamefont {T.~C.}\ \bibnamefont
  {Ralph}},\ }\href {\doibase 10.1103/PhysRevLett.114.060501} {\bibfield
  {journal} {\bibinfo  {journal} {Phys. Rev. Lett.}\ }\textbf {\bibinfo
  {volume} {114}},\ \bibinfo {pages} {060501} (\bibinfo {year}
  {2015})}\BibitemShut {NoStop}%
\bibitem [{\citenamefont {Hamilton}\ \emph {et~al.}(2017)\citenamefont
  {Hamilton}, \citenamefont {Kruse}, \citenamefont {Sansoni}, \citenamefont
  {Barkhofen}, \citenamefont {Silberhorn},\ and\ \citenamefont
  {Jex}}]{hamilton_gaussian_2017}%
  \BibitemOpen
  \bibfield  {author} {\bibinfo {author} {\bibfnamefont {C.~S.}\ \bibnamefont
  {Hamilton}}, \bibinfo {author} {\bibfnamefont {R.}~\bibnamefont {Kruse}},
  \bibinfo {author} {\bibfnamefont {L.}~\bibnamefont {Sansoni}}, \bibinfo
  {author} {\bibfnamefont {S.}~\bibnamefont {Barkhofen}}, \bibinfo {author}
  {\bibfnamefont {C.}~\bibnamefont {Silberhorn}}, \ and\ \bibinfo {author}
  {\bibfnamefont {I.}~\bibnamefont {Jex}},\ }\href {\doibase
  10.1103/PhysRevLett.119.170501} {\bibfield  {journal} {\bibinfo  {journal}
  {Phys. Rev. Lett.}\ }\textbf {\bibinfo {volume} {119}},\ \bibinfo {pages}
  {170501} (\bibinfo {year} {2017})}\BibitemShut {NoStop}%
\bibitem [{\citenamefont {Kruse}\ \emph {et~al.}(2019)\citenamefont {Kruse},
  \citenamefont {Hamilton}, \citenamefont {Sansoni}, \citenamefont {Barkhofen},
  \citenamefont {Silberhorn},\ and\ \citenamefont {Jex}}]{kruse_detailed_2019}%
  \BibitemOpen
  \bibfield  {author} {\bibinfo {author} {\bibfnamefont {R.}~\bibnamefont
  {Kruse}}, \bibinfo {author} {\bibfnamefont {C.~S.}\ \bibnamefont {Hamilton}},
  \bibinfo {author} {\bibfnamefont {L.}~\bibnamefont {Sansoni}}, \bibinfo
  {author} {\bibfnamefont {S.}~\bibnamefont {Barkhofen}}, \bibinfo {author}
  {\bibfnamefont {C.}~\bibnamefont {Silberhorn}}, \ and\ \bibinfo {author}
  {\bibfnamefont {I.}~\bibnamefont {Jex}},\ }\href {\doibase
  10.1103/PhysRevA.100.032326} {\bibfield  {journal} {\bibinfo  {journal}
  {Phys. Rev. A}\ }\textbf {\bibinfo {volume} {100}},\ \bibinfo {pages}
  {032326} (\bibinfo {year} {2019})}\BibitemShut {NoStop}%
\bibitem [{\citenamefont {Quesada}\ \emph {et~al.}(2018)\citenamefont
  {Quesada}, \citenamefont {Arrazola},\ and\ \citenamefont
  {Killoran}}]{quesada_gaussian_2018}%
  \BibitemOpen
  \bibfield  {author} {\bibinfo {author} {\bibfnamefont {N.}~\bibnamefont
  {Quesada}}, \bibinfo {author} {\bibfnamefont {J.~M.}\ \bibnamefont
  {Arrazola}}, \ and\ \bibinfo {author} {\bibfnamefont {N.}~\bibnamefont
  {Killoran}},\ }\href {\doibase 10.1103/PhysRevA.98.062322} {\bibfield
  {journal} {\bibinfo  {journal} {Phys. Rev. A}\ }\textbf {\bibinfo {volume}
  {98}},\ \bibinfo {pages} {062322} (\bibinfo {year} {2018})}\BibitemShut
  {NoStop}%
\bibitem [{\citenamefont {Madsen}\ \emph {et~al.}(2022)\citenamefont {Madsen},
  \citenamefont {Laudenbach}, \citenamefont {Askarani}, \citenamefont
  {Rortais}, \citenamefont {Vincent}, \citenamefont {Bulmer}, \citenamefont
  {Miatto}, \citenamefont {Neuhaus}, \citenamefont {Helt}, \citenamefont
  {Collins}, \citenamefont {Lita}, \citenamefont {Gerrits}, \citenamefont
  {Nam}, \citenamefont {Vaidya}, \citenamefont {Menotti}, \citenamefont
  {Dhand}, \citenamefont {Vernon}, \citenamefont {Quesada},\ and\ \citenamefont
  {Lavoie}}]{madsen_quantum_2022}%
  \BibitemOpen
  \bibfield  {author} {\bibinfo {author} {\bibfnamefont {L.~S.}\ \bibnamefont
  {Madsen}}, \bibinfo {author} {\bibfnamefont {F.}~\bibnamefont {Laudenbach}},
  \bibinfo {author} {\bibfnamefont {M.~F.}\ \bibnamefont {Askarani}}, \bibinfo
  {author} {\bibfnamefont {F.}~\bibnamefont {Rortais}}, \bibinfo {author}
  {\bibfnamefont {T.}~\bibnamefont {Vincent}}, \bibinfo {author} {\bibfnamefont
  {J.~F.~F.}\ \bibnamefont {Bulmer}}, \bibinfo {author} {\bibfnamefont {F.~M.}\
  \bibnamefont {Miatto}}, \bibinfo {author} {\bibfnamefont {L.}~\bibnamefont
  {Neuhaus}}, \bibinfo {author} {\bibfnamefont {L.~G.}\ \bibnamefont {Helt}},
  \bibinfo {author} {\bibfnamefont {M.~J.}\ \bibnamefont {Collins}}, \bibinfo
  {author} {\bibfnamefont {A.~E.}\ \bibnamefont {Lita}}, \bibinfo {author}
  {\bibfnamefont {T.}~\bibnamefont {Gerrits}}, \bibinfo {author} {\bibfnamefont
  {S.~W.}\ \bibnamefont {Nam}}, \bibinfo {author} {\bibfnamefont {V.~D.}\
  \bibnamefont {Vaidya}}, \bibinfo {author} {\bibfnamefont {M.}~\bibnamefont
  {Menotti}}, \bibinfo {author} {\bibfnamefont {I.}~\bibnamefont {Dhand}},
  \bibinfo {author} {\bibfnamefont {Z.}~\bibnamefont {Vernon}}, \bibinfo
  {author} {\bibfnamefont {N.}~\bibnamefont {Quesada}}, \ and\ \bibinfo
  {author} {\bibfnamefont {J.}~\bibnamefont {Lavoie}},\ }\href {\doibase
  10.1038/s41586-022-04725-x} {\bibfield  {journal} {\bibinfo  {journal}
  {Nature}\ }\textbf {\bibinfo {volume} {606}},\ \bibinfo {pages} {75}
  (\bibinfo {year} {2022})}\BibitemShut {NoStop}%
\bibitem [{\citenamefont {Aaronson}\ and\ \citenamefont
  {Arkhipov}(2010)}]{aaronson_computational_2010}%
  \BibitemOpen
  \bibfield  {author} {\bibinfo {author} {\bibfnamefont {S.}~\bibnamefont
  {Aaronson}}\ and\ \bibinfo {author} {\bibfnamefont {A.}~\bibnamefont
  {Arkhipov}},\ }\href {http://arxiv.org/abs/1011.3245} {\bibfield  {journal}
  {\bibinfo  {journal} {arXiv:1011.3245 [quant-ph]}\ } (\bibinfo {year}
  {2010})}\BibitemShut {NoStop}%
\bibitem [{\citenamefont {Huh}\ and\ \citenamefont
  {Yung}(2017)}]{huh_vibronic_2017}%
  \BibitemOpen
  \bibfield  {author} {\bibinfo {author} {\bibfnamefont {J.}~\bibnamefont
  {Huh}}\ and\ \bibinfo {author} {\bibfnamefont {M.-H.}\ \bibnamefont {Yung}},\
  }\href {\doibase 10.1038/s41598-017-07770-z} {\bibfield  {journal} {\bibinfo
  {journal} {Scientific Reports}\ }\textbf {\bibinfo {volume} {7}},\ \bibinfo
  {pages} {7462} (\bibinfo {year} {2017})}\BibitemShut {NoStop}%
\bibitem [{\citenamefont {Banchi}\ \emph {et~al.}(2020)\citenamefont {Banchi},
  \citenamefont {Fingerhuth}, \citenamefont {Babej}, \citenamefont {Ing},\ and\
  \citenamefont {Arrazola}}]{banchi_molecular_2020}%
  \BibitemOpen
  \bibfield  {author} {\bibinfo {author} {\bibfnamefont {L.}~\bibnamefont
  {Banchi}}, \bibinfo {author} {\bibfnamefont {M.}~\bibnamefont {Fingerhuth}},
  \bibinfo {author} {\bibfnamefont {T.}~\bibnamefont {Babej}}, \bibinfo
  {author} {\bibfnamefont {C.}~\bibnamefont {Ing}}, \ and\ \bibinfo {author}
  {\bibfnamefont {J.~M.}\ \bibnamefont {Arrazola}},\ }\href {\doibase
  10.1126/sciadv.aax1950} {\bibfield  {journal} {\bibinfo  {journal} {Sci.
  Adv.}\ }\textbf {\bibinfo {volume} {6}},\ \bibinfo {pages} {eaax1950}
  (\bibinfo {year} {2020})}\BibitemShut {NoStop}%
\bibitem [{\citenamefont {Bradler}\ \emph {et~al.}(2018)\citenamefont
  {Bradler}, \citenamefont {Dallaire-Demers}, \citenamefont {Rebentrost},
  \citenamefont {Su},\ and\ \citenamefont {Weedbrook}}]{bradler_gaussian_2018}%
  \BibitemOpen
  \bibfield  {author} {\bibinfo {author} {\bibfnamefont {K.}~\bibnamefont
  {Bradler}}, \bibinfo {author} {\bibfnamefont {P.-L.}\ \bibnamefont
  {Dallaire-Demers}}, \bibinfo {author} {\bibfnamefont {P.}~\bibnamefont
  {Rebentrost}}, \bibinfo {author} {\bibfnamefont {D.}~\bibnamefont {Su}}, \
  and\ \bibinfo {author} {\bibfnamefont {C.}~\bibnamefont {Weedbrook}},\ }\href
  {\doibase 10.1103/PhysRevA.98.032310} {\bibfield  {journal} {\bibinfo
  {journal} {Phys. Rev. A}\ }\textbf {\bibinfo {volume} {98}},\ \bibinfo
  {pages} {032310} (\bibinfo {year} {2018})}\BibitemShut {NoStop}%
\bibitem [{\citenamefont {Arrazola}\ and\ \citenamefont
  {Bromley}(2018)}]{arrazola_using_2018}%
  \BibitemOpen
  \bibfield  {author} {\bibinfo {author} {\bibfnamefont {J.~M.}\ \bibnamefont
  {Arrazola}}\ and\ \bibinfo {author} {\bibfnamefont {T.~R.}\ \bibnamefont
  {Bromley}},\ }\href {\doibase 10.1103/PhysRevLett.121.030503} {\bibfield
  {journal} {\bibinfo  {journal} {Phys. Rev. Lett.}\ }\textbf {\bibinfo
  {volume} {121}},\ \bibinfo {pages} {030503} (\bibinfo {year}
  {2018})}\BibitemShut {NoStop}%
\bibitem [{\citenamefont {Brádler}\ \emph {et~al.}(2021)\citenamefont
  {Brádler}, \citenamefont {Friedland}, \citenamefont {Izaac}, \citenamefont
  {Killoran},\ and\ \citenamefont {Su}}]{bradler_graph_2021}%
  \BibitemOpen
  \bibfield  {author} {\bibinfo {author} {\bibfnamefont {K.}~\bibnamefont
  {Brádler}}, \bibinfo {author} {\bibfnamefont {S.}~\bibnamefont {Friedland}},
  \bibinfo {author} {\bibfnamefont {J.}~\bibnamefont {Izaac}}, \bibinfo
  {author} {\bibfnamefont {N.}~\bibnamefont {Killoran}}, \ and\ \bibinfo
  {author} {\bibfnamefont {D.}~\bibnamefont {Su}},\ }\href {\doibase
  10.1515/spma-2020-0132} {\bibfield  {journal} {\bibinfo  {journal} {Special
  Matrices}\ }\textbf {\bibinfo {volume} {9}},\ \bibinfo {pages} {166}
  (\bibinfo {year} {2021})}\BibitemShut {NoStop}%
\bibitem [{\citenamefont {Arrazola}\ \emph {et~al.}(2018)\citenamefont
  {Arrazola}, \citenamefont {Bromley},\ and\ \citenamefont
  {Rebentrost}}]{arrazola_quantum_2018}%
  \BibitemOpen
  \bibfield  {author} {\bibinfo {author} {\bibfnamefont {J.~M.}\ \bibnamefont
  {Arrazola}}, \bibinfo {author} {\bibfnamefont {T.~R.}\ \bibnamefont
  {Bromley}}, \ and\ \bibinfo {author} {\bibfnamefont {P.}~\bibnamefont
  {Rebentrost}},\ }\href {\doibase 10.1103/PhysRevA.98.012322} {\bibfield
  {journal} {\bibinfo  {journal} {Phys. Rev. A}\ }\textbf {\bibinfo {volume}
  {98}},\ \bibinfo {pages} {012322} (\bibinfo {year} {2018})}\BibitemShut
  {NoStop}%
\bibitem [{\citenamefont {Schuld}\ and\ \citenamefont
  {Killoran}(2019)}]{schuld_quantum_2019}%
  \BibitemOpen
  \bibfield  {author} {\bibinfo {author} {\bibfnamefont {M.}~\bibnamefont
  {Schuld}}\ and\ \bibinfo {author} {\bibfnamefont {N.}~\bibnamefont
  {Killoran}},\ }\href {\doibase 10.1103/PhysRevLett.122.040504} {\bibfield
  {journal} {\bibinfo  {journal} {Phys. Rev. Lett.}\ }\textbf {\bibinfo
  {volume} {122}},\ \bibinfo {pages} {040504} (\bibinfo {year}
  {2019})}\BibitemShut {NoStop}%
\bibitem [{\citenamefont {Su}\ \emph {et~al.}(2021)\citenamefont {Su},
  \citenamefont {Israel}, \citenamefont {Sharma}, \citenamefont {Qi},
  \citenamefont {Dhand},\ and\ \citenamefont {Brádler}}]{su_error_2021}%
  \BibitemOpen
  \bibfield  {author} {\bibinfo {author} {\bibfnamefont {D.}~\bibnamefont
  {Su}}, \bibinfo {author} {\bibfnamefont {R.}~\bibnamefont {Israel}}, \bibinfo
  {author} {\bibfnamefont {K.}~\bibnamefont {Sharma}}, \bibinfo {author}
  {\bibfnamefont {H.}~\bibnamefont {Qi}}, \bibinfo {author} {\bibfnamefont
  {I.}~\bibnamefont {Dhand}}, \ and\ \bibinfo {author} {\bibfnamefont
  {K.}~\bibnamefont {Brádler}},\ }\href {\doibase 10.22331/q-2021-05-04-452}
  {\bibfield  {journal} {\bibinfo  {journal} {Quantum}\ }\textbf {\bibinfo
  {volume} {5}},\ \bibinfo {pages} {452} (\bibinfo {year} {2021})}\BibitemShut
  {NoStop}%
\bibitem [{\citenamefont {Qi}\ \emph {et~al.}(2020)\citenamefont {Qi},
  \citenamefont {Brod}, \citenamefont {Quesada},\ and\ \citenamefont
  {García-Patrón}}]{qi_regimes_2020}%
  \BibitemOpen
  \bibfield  {author} {\bibinfo {author} {\bibfnamefont {H.}~\bibnamefont
  {Qi}}, \bibinfo {author} {\bibfnamefont {D.~J.}\ \bibnamefont {Brod}},
  \bibinfo {author} {\bibfnamefont {N.}~\bibnamefont {Quesada}}, \ and\
  \bibinfo {author} {\bibfnamefont {R.}~\bibnamefont {García-Patrón}},\
  }\href {\doibase 10.1103/PhysRevLett.124.100502} {\bibfield  {journal}
  {\bibinfo  {journal} {Phys. Rev. Lett.}\ }\textbf {\bibinfo {volume} {124}},\
  \bibinfo {pages} {100502} (\bibinfo {year} {2020})}\BibitemShut {NoStop}%
\bibitem [{\citenamefont {Oh}\ \emph {et~al.}(2022)\citenamefont {Oh},
  \citenamefont {Lim}, \citenamefont {Fefferman},\ and\ \citenamefont
  {Jiang}}]{oh_classical_2022}%
  \BibitemOpen
  \bibfield  {author} {\bibinfo {author} {\bibfnamefont {C.}~\bibnamefont
  {Oh}}, \bibinfo {author} {\bibfnamefont {Y.}~\bibnamefont {Lim}}, \bibinfo
  {author} {\bibfnamefont {B.}~\bibnamefont {Fefferman}}, \ and\ \bibinfo
  {author} {\bibfnamefont {L.}~\bibnamefont {Jiang}},\ }\href
  {https://link.aps.org/doi/10.1103/PhysRevLett.128.190501} {\bibfield
  {journal} {\bibinfo  {journal} {Phys. Rev. Lett.}\ }\textbf {\bibinfo
  {volume} {128}} (\bibinfo {year} {2022})}\BibitemShut {NoStop}%
\bibitem [{\citenamefont {Wang}\ \emph {et~al.}(2007)\citenamefont {Wang},
  \citenamefont {Hiroshima}, \citenamefont {Tomita},\ and\ \citenamefont
  {Hayashi}}]{wang_quantum_2007}%
  \BibitemOpen
  \bibfield  {author} {\bibinfo {author} {\bibfnamefont {X.}~\bibnamefont
  {Wang}}, \bibinfo {author} {\bibfnamefont {T.}~\bibnamefont {Hiroshima}},
  \bibinfo {author} {\bibfnamefont {A.}~\bibnamefont {Tomita}}, \ and\ \bibinfo
  {author} {\bibfnamefont {M.}~\bibnamefont {Hayashi}},\ }\href {\doibase
  10.1016/j.physrep.2007.04.005} {\bibfield  {journal} {\bibinfo  {journal}
  {Physics Reports}\ }\textbf {\bibinfo {volume} {448}},\ \bibinfo {pages} {1}
  (\bibinfo {year} {2007})}\BibitemShut {NoStop}%
\bibitem [{\citenamefont {Serafini}(2017)}]{serafini_quantum_2017}%
  \BibitemOpen
  \bibfield  {author} {\bibinfo {author} {\bibfnamefont {A.}~\bibnamefont
  {Serafini}},\ }\href@noop {} {\emph {\bibinfo {title} {Quantum continuous
  variables: a primer of theoretical methods}}}\ (\bibinfo  {publisher} {CRC
  Press, Taylor \& Francis Group, CRC Press is an imprint of the Taylor \&
  Francis Group, an informa business},\ \bibinfo {address} {Boca Raton},\
  \bibinfo {year} {2017})\BibitemShut {NoStop}%
\bibitem [{\citenamefont {Barnett}\ and\ \citenamefont
  {Radmore}(2002)}]{barnett_methods_2002}%
  \BibitemOpen
  \bibfield  {author} {\bibinfo {author} {\bibfnamefont {S.}~\bibnamefont
  {Barnett}}\ and\ \bibinfo {author} {\bibfnamefont {P.}~\bibnamefont
  {Radmore}},\ }\href {\doibase 10.1093/acprof:oso/9780198563617.001.0001}
  {\emph {\bibinfo {title} {Methods in {Theoretical} {Quantum} {Optics}}}}\
  (\bibinfo  {publisher} {Oxford University Press},\ \bibinfo {year}
  {2002})\BibitemShut {NoStop}%
\bibitem [{\citenamefont {Scully}\ and\ \citenamefont
  {Zubairy}(1997)}]{scully_quantum_1997}%
  \BibitemOpen
  \bibfield  {author} {\bibinfo {author} {\bibfnamefont {M.~O.}\ \bibnamefont
  {Scully}}\ and\ \bibinfo {author} {\bibfnamefont {M.~S.}\ \bibnamefont
  {Zubairy}},\ }\href {\doibase 10.1017/CBO9780511813993} {\emph {\bibinfo
  {title} {Quantum {Optics}}}},\ \bibinfo {edition} {1st}\ ed.\ (\bibinfo
  {publisher} {Cambridge University Press},\ \bibinfo {year}
  {1997})\BibitemShut {NoStop}%
\bibitem [{\citenamefont {Rohatgi}\ and\ \citenamefont
  {Saleh}(2015)}]{rohatgi_introduction_2015}%
  \BibitemOpen
  \bibfield  {author} {\bibinfo {author} {\bibfnamefont {V.~K.}\ \bibnamefont
  {Rohatgi}}\ and\ \bibinfo {author} {\bibfnamefont {A.~K. M.~E.}\ \bibnamefont
  {Saleh}},\ }\href@noop {} {\emph {\bibinfo {title} {An introduction to
  probability and statistics}}},\ \bibinfo {edition} {third edition}\ ed.,\
  Wiley series in probability and statistics\ (\bibinfo  {publisher} {Wiley},\
  \bibinfo {address} {Hoboken, New Jersey},\ \bibinfo {year}
  {2015})\BibitemShut {NoStop}%
\bibitem [{\citenamefont {Brod}\ and\ \citenamefont
  {Oszmaniec}(2020)}]{brod_classical_2020}%
  \BibitemOpen
  \bibfield  {author} {\bibinfo {author} {\bibfnamefont {D.~J.}\ \bibnamefont
  {Brod}}\ and\ \bibinfo {author} {\bibfnamefont {M.}~\bibnamefont
  {Oszmaniec}},\ }\href {\doibase 10.22331/q-2020-05-14-267} {\bibfield
  {journal} {\bibinfo  {journal} {Quantum}\ }\textbf {\bibinfo {volume} {4}},\
  \bibinfo {pages} {267} (\bibinfo {year} {2020})}\BibitemShut {NoStop}%
\bibitem [{\citenamefont {García-Patrón}\ \emph {et~al.}(2019)\citenamefont
  {García-Patrón}, \citenamefont {Renema},\ and\ \citenamefont
  {Shchesnovich}}]{garcia-patron_simulating_2019}%
  \BibitemOpen
  \bibfield  {author} {\bibinfo {author} {\bibfnamefont {R.}~\bibnamefont
  {García-Patrón}}, \bibinfo {author} {\bibfnamefont {J.~J.}\ \bibnamefont
  {Renema}}, \ and\ \bibinfo {author} {\bibfnamefont {V.}~\bibnamefont
  {Shchesnovich}},\ }\href {\doibase 10.22331/q-2019-08-05-169} {\bibfield
  {journal} {\bibinfo  {journal} {Quantum}\ }\textbf {\bibinfo {volume} {3}},\
  \bibinfo {pages} {169} (\bibinfo {year} {2019})}\BibitemShut {NoStop}%
\bibitem [{\citenamefont {Oszmaniec}\ and\ \citenamefont
  {Brod}(2018)}]{oszmaniec_classical_2018}%
  \BibitemOpen
  \bibfield  {author} {\bibinfo {author} {\bibfnamefont {M.}~\bibnamefont
  {Oszmaniec}}\ and\ \bibinfo {author} {\bibfnamefont {D.~J.}\ \bibnamefont
  {Brod}},\ }\href {\doibase 10.1088/1367-2630/aadfa8} {\bibfield  {journal}
  {\bibinfo  {journal} {New J. Phys.}\ }\textbf {\bibinfo {volume} {20}},\
  \bibinfo {pages} {092002} (\bibinfo {year} {2018})}\BibitemShut {NoStop}%
\bibitem [{\citenamefont {Quesada}\ and\ \citenamefont
  {Arrazola}(2020)}]{quesada_exact_2020}%
  \BibitemOpen
  \bibfield  {author} {\bibinfo {author} {\bibfnamefont {N.}~\bibnamefont
  {Quesada}}\ and\ \bibinfo {author} {\bibfnamefont {J.~M.}\ \bibnamefont
  {Arrazola}},\ }\href {\doibase 10.1103/PhysRevResearch.2.023005} {\bibfield
  {journal} {\bibinfo  {journal} {Phys. Rev. Research}\ }\textbf {\bibinfo
  {volume} {2}},\ \bibinfo {pages} {023005} (\bibinfo {year}
  {2020})}\BibitemShut {NoStop}%
\bibitem [{\citenamefont {Villalonga}\ \emph {et~al.}(2022)\citenamefont
  {Villalonga}, \citenamefont {Niu}, \citenamefont {Li}, \citenamefont {Neven},
  \citenamefont {Platt}, \citenamefont {Smelyanskiy},\ and\ \citenamefont
  {Boixo}}]{villalonga_efficient_2022}%
  \BibitemOpen
  \bibfield  {author} {\bibinfo {author} {\bibfnamefont {B.}~\bibnamefont
  {Villalonga}}, \bibinfo {author} {\bibfnamefont {M.~Y.}\ \bibnamefont {Niu}},
  \bibinfo {author} {\bibfnamefont {L.}~\bibnamefont {Li}}, \bibinfo {author}
  {\bibfnamefont {H.}~\bibnamefont {Neven}}, \bibinfo {author} {\bibfnamefont
  {J.~C.}\ \bibnamefont {Platt}}, \bibinfo {author} {\bibfnamefont {V.~N.}\
  \bibnamefont {Smelyanskiy}}, \ and\ \bibinfo {author} {\bibfnamefont
  {S.}~\bibnamefont {Boixo}},\ }\href {http://arxiv.org/abs/2109.11525}
  {\bibfield  {journal} {\bibinfo  {journal} {arXiv:2109.11525 [quant-ph]}\ }
  (\bibinfo {year} {2022})}\BibitemShut {NoStop}%
\bibitem [{\citenamefont {Martínez-Cifuentes}\ \emph
  {et~al.}(2022)\citenamefont {Martínez-Cifuentes}, \citenamefont
  {Fonseca-Romero},\ and\ \citenamefont
  {Quesada}}]{martinez-cifuentes_classical_2022}%
  \BibitemOpen
  \bibfield  {author} {\bibinfo {author} {\bibfnamefont {J.}~\bibnamefont
  {Martínez-Cifuentes}}, \bibinfo {author} {\bibfnamefont {K.~M.}\
  \bibnamefont {Fonseca-Romero}}, \ and\ \bibinfo {author} {\bibfnamefont
  {N.}~\bibnamefont {Quesada}},\ }\href {http://arxiv.org/abs/2207.10058}
  {\bibfield  {journal} {\bibinfo  {journal} {arXiv:2207.10058 [quant-ph]}\ }
  (\bibinfo {year} {2022})}\BibitemShut {NoStop}%
\bibitem [{\citenamefont {Shi}\ and\ \citenamefont
  {Byrnes}(2022)}]{shi_effect_2022}%
  \BibitemOpen
  \bibfield  {author} {\bibinfo {author} {\bibfnamefont {J.}~\bibnamefont
  {Shi}}\ and\ \bibinfo {author} {\bibfnamefont {T.}~\bibnamefont {Byrnes}},\
  }\href {\doibase 10.1038/s41534-022-00557-9} {\bibfield  {journal} {\bibinfo
  {journal} {npj Quantum Information}\ }\textbf {\bibinfo {volume} {8}},\
  \bibinfo {pages} {54} (\bibinfo {year} {2022})}\BibitemShut {NoStop}%
\bibitem [{\citenamefont {Rohde}\ \emph {et~al.}(2015)\citenamefont {Rohde},
  \citenamefont {Motes}, \citenamefont {Knott}, \citenamefont {Fitzsimons},
  \citenamefont {Munro},\ and\ \citenamefont {Dowling}}]{rohde_evidence_2015}%
  \BibitemOpen
  \bibfield  {author} {\bibinfo {author} {\bibfnamefont {P.~P.}\ \bibnamefont
  {Rohde}}, \bibinfo {author} {\bibfnamefont {K.~R.}\ \bibnamefont {Motes}},
  \bibinfo {author} {\bibfnamefont {P.~A.}\ \bibnamefont {Knott}}, \bibinfo
  {author} {\bibfnamefont {J.}~\bibnamefont {Fitzsimons}}, \bibinfo {author}
  {\bibfnamefont {W.~J.}\ \bibnamefont {Munro}}, \ and\ \bibinfo {author}
  {\bibfnamefont {J.~P.}\ \bibnamefont {Dowling}},\ }\href {\doibase
  10.1103/PhysRevA.91.012342} {\bibfield  {journal} {\bibinfo  {journal} {Phys.
  Rev. A}\ }\textbf {\bibinfo {volume} {91}},\ \bibinfo {pages} {012342}
  (\bibinfo {year} {2015})}\BibitemShut {NoStop}%
\end{thebibliography}%


%
\end{document}